\documentclass[11pt, a4paper, oneside]{Thesis} 

\usepackage[english]{babel}

\usepackage{multirow}
\usepackage{url}

\usepackage[tablename=Table]{caption}
\usepackage{amsmath}
\usepackage{amsthm}

\usepackage[ruled,vlined,linesnumbered]{algorithm2e}
\SetAlFnt{\small}
\SetAlCapFnt{\small}
\SetAlCapNameFnt{\small}
\usepackage{algorithmic}
\algsetup{linenosize=\large}

\usepackage{footmisc}
\usepackage{diagbox}
\usepackage{footmisc}
\usepackage{pgfplots}
\pgfplotsset{compat=1.8}
\usepackage{tikz}
\usepackage{float}
\usepackage{amsfonts}
\usepackage[normalem]{ulem}
\usepackage{booktabs}
\usepackage{multirow}
\usepackage{array,arydshln,xcolor}

\newcommand{\ra}[1]{\renewcommand{\arraystretch}{#1}}

\usetikzlibrary{arrows,shapes,positioning,shadows,trees}

\graphicspath{{./Pictures/}} 
\usepackage{graphicx}
\usepackage{subfigure}
\usepackage{pdfpages}
\usepackage{changepage}
\usetikzlibrary{calc}
\usetikzlibrary{decorations.pathmorphing}

\usepackage[square, numbers, comma, sort]{natbib} 
\usepackage[utf8]{inputenc}
\hypersetup{urlcolor=black, colorlinks=false} 
\title{\ttitle} 

\begin{document}
\renewcommand{\listtablename}{List of tables}
\renewcommand{\bibname}{References}
\frontmatter 
\setstretch{1.3} 

\fancyhead{} 
\rhead{\thepage} 
\lhead{} 

\pagestyle{fancy} 

\newcommand{\HRule}{\rule{\linewidth}{0.5mm}} 

\hypersetup{pdftitle={\ttitle}}
\hypersetup{pdfsubject=\subjectname}
\hypersetup{pdfauthor=\authornames}
\hypersetup{pdfkeywords=\keywordnames}


\begin{titlepage}
\begin{adjustwidth}{-0.9cm}{0.5cm}
\begin{center}
\begin{tikzpicture}[overlay,remember picture]
        ($ (current page.north west) + (4cm,1cm) $)
        rectangle
        ($ (current page.south east) + (1cm,4cm) $);
\end{tikzpicture}

\includegraphics[scale=0.75]{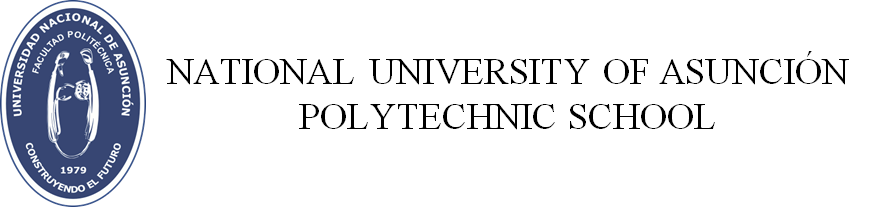} 

\HRule \\[0.4cm] 
{\huge
    \textsc{ 
        Scalarization Methods for Many-Objective Virtual Machine Placement of Elastic Infrastructures in Overbooked Cloud Computing Data Centers Under Uncertainty 
    }
}\\[0.2cm] 
\HRule \\[1.5cm] 
 
\textsc{
    \LARGE { Informatics Engineering Final Work }
}\\[0.3cm]
 
\vfill
\textsc{\LARGE César Augusto Amarilla Cardozo}\\[1.5cm]

\vfill

\textsc{\LARGE San Lorenzo - Paraguay}\\[1.5cm]
\textsc{\LARGE 2017}\\[1.5cm]
 
\vfill
\end{center}
\end{adjustwidth}
\end{titlepage}

\clearpage


\begin{titlepage}
\begin{center}
\includegraphics[scale=0.75]{./Figures/logo_title.png} 

\HRule \\[0.4cm] 
{\huge \bfseries Scalarization Methods for Many-Objective Virtual Machine Placement of Elastic Infrastructures in Overbooked Cloud Computing Data Centers Under Uncertainty }\\[0.4cm] 
\HRule \\[1.5cm] 
 
\large \textit{ Informatics Engineering Final Work }\\[0.3cm]
 
 \vfill
 
\begin{minipage}{0.6\textwidth}
\begin{center} \large
\emph{Author:}\\
{César Augusto Amarilla Cardozo}
\end{center}
\end{minipage}

\vfill

\begin{minipage}{0.4\textwidth}
\begin{center} \large
\emph{Advisors:} \\
{D.Sc. Fabio López-Pires\\}
{D.Sc. Benjamín Barán} \\
\end{center}
\end{minipage}\\[3cm]
 
\textsc{\LARGE San Lorenzo - Paraguay}\\[1.5cm]
\textsc{\LARGE 2017}\\[1.5cm]
 
\vfill
\end{center}
\end{titlepage}

\hyphenation{va-ria-bi-li-ty pro-blem stu-died off-line sol-ving pro-vi-sio-ning con-si-de-ring cons-traints phy-si-cal Ex-pe-ri-ments heu-ris-tics op-ti-mi-zing up-co-ming con-si-de-red si-mi-lar-ly o-pe-ra-tors e-vo-lu-tio-na-ry a-vai-la-ble fo-llo-wing ge-ne-ra-ted ta-king pro-blems o-ver-boo-king fe-de-ra-tion con-si-de-ra-tions Pro-ba-bi-li-ty e-qu-a-lly a-ve-ra-ge com-pa-ri-son fin-dings nor-ma-li-zed}

\clearpage 
\setstretch{1.3} 
\pagestyle{empty} 
\dedicatory{
    \begin{flushright}
        I'm the red one
        
        We all have strange nicknames
        
        The big one, the blonde one, the brown one and the young one
        
        A strange bunch of people I know, they are noisy and full of joy
        
        I know them from quite a long time now
        
        Never had the distaste of missing them
        
        They are there all the time
        
        When I need them
    \end{flushright}

} 
\addtocontents{toc}{\vspace{2em}} 

\clearpage 
\setstretch{1.3} 
\acknowledgements{
    \addtocontents{toc}{\vspace{1em}} 
    
    The final grade project is the last step on a long walk that changed the world for me in so many ways. It started as a logical continuation of a path chosen in my teen years but it gradually grew into a life-time passion. Professors and fellow students helped throughout the whole class, terms and exams process of the college career. All of them deserve to be thanked for the support to even become eligible for a final grade project.
    The last two years I worked under the supervision of my two supervisors Professors Benjamín Barán and Fabio López-Pires, who have transmitted to me their knowledge and experiences with such patience and diligence. Their advice and guidance were irreplaceable to complete this work.
    
    I would like to thank my research colleagues, who at times provided the needed push to reach the goals but also provided advice on when to leave the work and enjoy the little thins. Many thanks to Saúl Zalimben, Leonardo Benitez and Rodrigo Ferreira.
    
    The National University of Asunción and the Scientific and Applied Computing Laboratory (LCCA) where very helpful to provide the needed installations to carry out the experiments needed and to support the work that we produced in order to reach other continents, universities and colleagues.
    
    At last, I want to thank the pillar of this effort, the people who saw the remains of long days of hard work and welcomed my complaints about life itself when things got very difficult. They have provided the essentials to keep me going, a laugh at the table, a smile at the morning, a warm dinner at cold nights and nice stories to distract the mind. The family that I got is amazing and should be thanked at last because they deserve to be the final and grand brooch of this thanks note. I wouldn't have done it without them.

}




\clearpage 
\abstract{\addtocontents{toc}{\vspace{1em}}} 

Cloud computing datacenters provide thousands to millions of virtual machines (VMs) on-demand in highly dynamic environments, requiring quick placement of requested VMs into available physical machines (PMs). Due to the randomness of customer requests, the Virtual Machine Placement (VMP) should be formulated as an online optimization problem. 

The first part of this work proposes a formulation of a provider-oriented VMP problem for federated-cloud considering the optimization of the following objective functions: (i) power consumption, (ii) economical revenue, (iii) quality of service and (iv) resource utilization. In order to analyze alternatives to solve the formulated problem, an experimental comparison of five different online deterministic heuristics against an offline memetic algorithm with migration of VMs was performed, considering several experimental workloads. Simulations indicate that First-Fit Decreasing algorithm (A4) outperforms other evaluated heuristics on average. Experimental results proved that an offline memetic algorithm improves the quality of the solutions with migrations of VMs at the expense of placement reconfigurations.

The second part of this work presents a two-phase schema formulation of a VMP problem considering the optimization of three objective 
functions: (i) power consumption, (ii) economical revenue, and (iii) resource utilization, in an IaaS environment with elasticity and overbooking capabilities. The two-phase schema formulation describes that the allocation of the VMs can be separated into two sub-problems, the incremental allocation (iVMP) and the reconfiguration of a placement (VMPr). In order to portray the dynamic nature of an IaaS environment a customizable workload trace generator was developed to simulate uncertainty in the scenarios with elasticity and overbooking of resources in VM requests.
To analyze alternatives to solve the formulated problem, an experimental comparison of three different objective function scalarization methods as part of the iVMP and VMPr was performed considering several experimental workloads. 
Simulations indicate that the Euclidean distance to origin outperforms other evaluated scalarization methods on average.
Experimental results proved that the Euclidean distance is preferable over the other scalarizatiom methods to improve the values of the power consumption objective function.


\clearpage 
\pagestyle{fancy}
\lhead{\emph{Content}} 


\setcounter{tocdepth}{5} 
\tableofcontents 


\lhead{\emph{List of Figures}}  
\listoffigures            
\lhead{\emph{List of Tables}}   
\listoftables                   

\clearpage 
\setstretch{1.5} 
\lhead{\emph{Acronyms and symbols}} 
\listofsymbols{ll} 
{

\textbf{ANSI} & \textit{American National Standards Institute}\\
\textbf{AWS} & \textit{Amazon Web Services}\\
\textbf{BF} & \textit{Best Fit}\\
\textbf{BFD} & \textit{Best Fit Decreasing}\\
\textbf{CD} & \textit{Chebyshev Distance}\\
\textbf{CPU} & \textit{Central Processing Unit}\\
\textbf{CSV} & \textit{Comma Separated Values}\\
\textbf{CWTG} & \textit{Cloud Workload Trace Generator}\\
\textbf{DRF} & \textit{Dominant Resource Fit}\\
\textbf{EC2} & \textit{Elastic Compute Cloud}\\
\textbf{ECU} & \textit{E2 Compute Unit}\\
\textbf{ED} & \textit{Euclidean Distance}\\
\textbf{FF} & \textit{First Fit}\\
\textbf{FFD} & \textit{First Fit Decreasing}\\
\textbf{GB} & \textit{Giba Byte}\\
\textbf{GHz} & \textit{Giga Hertz}\\
\textbf{GPU} & \textit{Graphics Processing Unit}\\
\textbf{IaaS} & \textit{Infraestructure as a Service}\\
\textbf{ILP} & \textit{Integer Linear Programming}\\
\textbf{iVMP} & \textit{Incremental VMP}\\
\textbf{L} & \textit{Large Instance Type}\\
\textbf{M} & \textit{Medium Instance Type}\\
\textbf{MA} & \textit{Memethic Algorithm}\\
\textbf{MAM} & \textit{Many-objective as Mono-objective}\\
\textbf{Mbps} & \textit{Mega bit per second}\\
\textbf{MF} & \textit{Main Finding}\\
\textbf{OF} & \textit{Objective Function}\\
\textbf{PaaS} & \textit{Platform as a Service}\\
\textbf{PDF} & \textit{Probability Distribution Function}\\
\textbf{PM} & \textit{Physical Machine}\\
\textbf{QoS} & \textit{Quality of Service}\\
\textbf{R} & \textit{Economical Revenue}\\
\textbf{RAM} & \textit{Random Access Memory}\\
\textbf{S} & \textit{Small Instance Type}\\
\textbf{SaaS} & \textit{Software as a Service}\\
\textbf{SLA} & \textit{Service Level Agreement}\\
\textbf{SM} & \textit{Scalarization Methods}\\
\textbf{USD} & \textit{United States' Dollar}\\
\textbf{VM} & \textit{Virtual Machine}\\
\textbf{VMP} & \textit{Virtual Machine Placement}\\
\textbf{VMPr} & \textit{VMP reconfiguration}\\
\textbf{W} & \textit{Watts}\\
\textbf{WF} & \textit{Worst Fit}\\
\textbf{WS} & \textit{Weighted Sum}\\
\textbf{XL} & \textit{Extra Large Instance Type}\\

}

\mainmatter 
\pagestyle{fancy} 

\clearpage 

\lhead{\emph{Introduction}}

\chapter{Introduction}

A main concern of cloud datacenters design is to efficiently manage available resources in order to improve performance and reduce energy consumption of a given computational infrastructure. Most of the time, servers operate in a very low energy-efficiency region (i.e. between $10\%$ and $50\%$ of resource utilization), even considering that workload peaks rarely occur in practice \cite{barroso2007case}. Several methods were considered for addressing this issue, being the virtualization of resources the most studied for cloud computing datacenters.

In cloud computing datacenters, resources are dynamically allocated and released in order to serve requested demands. In this context, deciding the right allocation of virtual machines (VMs) into physical machines (PMs) is known in the specialized literature as Virtual Machine Placement (VMP).

Cloud computing datacenters deliver infrastructure (IaaS), platform (PaaS) and software (SaaS) as services available to end users in a pay-as-you-go basis \cite{Mell2011}. Particularly, the Infrastructure as a Service model provides processing, storage, network and other fundamental computing resources where a customer is able to deploy and run arbitrary software, which can include operating systems and applications \cite{buyya2008market}.

Resources in cloud computing datacenters are viewed by their users as unlimited and always available assets that they can dispose of at will. Achieving this performance over limited resources is a central focus of research in the field. The principal consideration in this vein of research is to provide efficient solutions for the VMP problem in order to take the most advantage of the available resources. Consequently, a main concern of cloud computing datacenter design is to efficiently manage available resources in order to improve performance and reduce energy consumption of a given computational infrastructure.

It is relevant to remember that the VMP problem is a NP-Hard combinatorial optimization problem \cite{Speitkamp2010}. From an Infrastructure as a Service (IaaS) provider's perspective, the VMP problem must be formulated as an online problem (considering that requests from customers are unknown a priori) and  it should be solved with short time constraints achieving solutions with acceptable performance.

The first part of this work focuses in online and offline formulations of a provider-oriented VMP problem in federated-clouds \cite{lopez2015} considering the management of an extensible set of resources and the optimization of Objective Functions (OF). This formulation is proposed as a Many-objective as Mono-objective (MAM) formulation. As a mechanism of objective function values consolidation, a Weighted Sum Method is utilized.
Considering the proposed formulations, five of the most studied heuristics were compared against a memetic algorithm (MA) solving an offline formulation of the problem, while considering the optimization of four OF from the most studied objective functions groups according to \cite{lopez2015}. The main goal of the proposed experimental comparison is to analyze the online nature of the VMP problem optimizing many different objectives, identifying advantages and disadvantages of well-known online heuristics against offline alternatives such as the memetic algorithm (MA) proposed in \cite{Ihara2015f}.
Consequently, providing relevant information to design and implement resource-management systems, more specifically, resource allocation algorithms for VMP problems.

From an IaaS provider perspective, the VMP is mostly formulated as an online problem and must be solved with short time constraints \cite{lopez2015}. Online decisions made along the operation of a cloud computing infrastructure may negatively affect the quality of obtained solutions when compared to offline decisions \cite{lopez2016LANC}. Unfortunately, offline formulations are not appropriate for highly dynamic environments for real-world IaaS providers, where cloud services are requested dynamically. To improve the quality of solutions obtained by online algorithms, the VMP problem could be formulated as a two-phase optimization problem, combining advantages of online and offline formulations \cite{lopez2016LANC}. In this context, VMP problems could be decomposed in two different sub-problems: (i) incremental VMP (iVMP) and (ii) VMP reconfiguration (VMPr) (see Figure \ref{fig:two_phase_optimization_scheme}).

In order to model the highly dynamic environments considering IaaS provider's perspective, a workload trace generator was developed. The cloud environment workload trace generator (CWTG) takes into consideration IaaS providers environments that are based on elasticity and overbooking of physical resources. This considerations allow the generator to produce dynamic environments identified in a state-of-the-art taxonomy detailed in \cite{lopez2015}.

To model the complexity and dynamism of real time IaaS environments, the VMP formulation proposed as a MAM in the first part of this work is extended to support the uncertainty of a complex IaaS environments. Taking this into consideration a big part of the process of finding a possible placement of VMs is to calculate the benefits of such placement in order to compare the results of the intended allocation with each other. Scalarization methods must be used to consolidate the values of the different OF into a single comparable value, so that the performance of the comparison of solutions could improve.
 
Consequently, the second part of this work, based on the two-phase scheme formulation of a provider-oriented VMP for federated-cloud deployments, proposes an experimental evaluation under uncertainty of the three scalarization methods developed as part of the solution selection process of the iVMP phase in conjunction with a MA as part of the VMPr phase, simultaneously optimizing the following objective functions: (1) power consumption, (2) economical revenue, (3) resource utilization and (4) placement reconfiguration time, while considering elasticity and overbooking, using generated dynamic cloud workload traces.

\section{Thesis Objectives}

To deal with the research challenges associated to provider-oriented VMP problems above mentioned, \textbf{the following objectives have been delineated:}

\textbf{Part I.}
\begin{itemize}
	\item \textbf{Objective 1:} Propose a formulation of a provider-oriented VMP problem for federated-cloud deployments, considering the optimization of the following objective functions: (1) power consumption, (2) economical revenue, (3) quality of service and (4) resource utilization.
	\item \textbf{Objective 2:} Perform an experimental comparison of five different online deterministic heuristics against an offline Memetic Algorithm (MA) with migration of VMs for the resolution of the proposed formulation.
\end{itemize}

\textbf{Part II.}
\begin{itemize}
	\item \textbf{Objective 3:} Develop a workload trace generator for VMP problems, capable of generating dynamic environments identified in the state-of-the-art, including IaaS environments that takes into account both service elasticity and overbooking of physical resources.
	\item \textbf{Objective 4:} Implement in a state-of-the-art two-phase optimization scheme for VMP problems, the following objective function scalarization methods: Weighted Sum (WS), Euclidean Distance to origin (ED) and Chebyshev Distance to origin (CD).
    \item \textbf{Objective 5:} Perform an experimental evaluation under uncertainty of the objective function scalarization methods implemented as part of Objective 4.
\end{itemize}

\clearpage
\section{Thesis Organization}

\textbf{The remainder of this work is organized as follows:} 

Part 1 is structured in the following way: Chapter \ref{chap:problem_formulation_first_part} 
summarizes the proposed provider-oriented VMP problem formulation with many objectives. 
Chapter \ref{chap:algorithms_first_part} describes the presented algorithms to solve 
the formulated problem, while Chapter \ref{chap:experimental_results_first_part} presents 
experimental workloads, obtained results and main findings of the comparison. 
Finally, conclusions and future work of the first part are left to 
Chapter \ref{chap:conclusions_and_future_works_first_part}.

Part 2 is structured as follows: Chapter \ref{chap:problem_formulation_second_part} summarizes the 
two-phase optimization scheme provider-oriented VMP problem formulation with many objectives, 
while considering services of elasticity and overbooking. Chapter \ref{chap:cwtg} details the proposal of an extendable workload trace generator and the uncertainty parameters it supports. Chapter \ref{chap:experimental_results_second_part} presents the evaluated scalarization methods, experimental environment, obtained results and main findings of the comparison. Finally, conclusions and future work of the second part are left to Chapter \ref{chap:conclusions_and_future_works_second_part}.



\vspace*{\fill}
\textbf{
    {\huge
        \begin{center}
        \part{An Experimental Comparison of Algorithms for Virtual Machine Placement Considering Many Objectives}
        \end{center}
    }
}
\vspace*{\fill}

\lhead{Part I. \emph{An Experimental Comparison of Algorithms for Virtual Machine Placement Considering Many Objectives}} 

\vspace*{\fill}{
    \begin{center}
        \textcolor{gray}{ \textit{This page is intentionally left blank. }}    
    \end{center}
}
\vspace*{\fill}

\chapter{Part I - Related Work and Motivation}
\label{chap:related_work}

Multiple articles in the provider-oriented VMP literature have studied the problem in both offline and online environments. The following chapter mentions the relevant contributions of the studied research work in the area as well as detailing the motivation of the presented work.

\section{VMP considering Many Objectives}

López-Pires and Barán recently proposed in \cite{Ihara2015f} and \cite{lopez2015b} an offline formulations of VMP problems considering many objectives, proposing novel memetic algorithms for solving the formulated problems. Considering the on-demand model of cloud computing with dynamic resource provisioning and dynamic workloads of cloud applications \cite{Mell2011}, the resolution of the VMP problem should be performed as fast as possible. Consequently, applying only offline formulations of the VMP problem with meta-heuristics as solution technique, may not be appropriate for these dynamic environments. Therefore, as previously mentioned, solution techniques with low computational complexity (e.g. heuristics) are studied intensively for solving online formulations of VMP problems \cite{lopez2015}. 

In this context, Fang et al. presented in \cite{fang2013power} a validation of a proposed power-aware algorithm against well-known heuristics such as: BF, FF and WF. Additionally, Jin et al. studied FFD and Dominant Resource First (DRF) algorithms \cite{Jin2012}, while Anand et al. evaluated in \cite{anand2013virtual} a proposed Integer Linear Programming (ILP) formulation and a FFD algorithm to attend common limitations of ILP algorithms for large instances of NP-Hard optimization problems. On the contrary, this work does not compare novel algorithms against well-known heuristics as presented in the above mentioned work.

Experimental results presented by Ihara et al. in \cite{Ihara2015f} recommend the combination of many objective functions into a single objective for IaaS environments, given the requirement of obtaining solutions in a short period of time.

According to L\'opez-Pires and Bar\'an in \cite{lopez2015}, the most studied heuristics in the VMP literature are: Best-Fit (BF), Best-Fit Decreasing (BFD), First-Fit (FF), First-Fit Decreasing (FFD) and Worst-Fit (WF) and over 60 different objectives have been proposed for VMP problems. The number of considered objective functions may rapidly increase once a complete understanding of the VMP problem is accomplished for practical problems, where many different parameters should be ideally taken into account. Consequently, a formulation of a VMP problem is presented, considering the optimization of the following four objective functions: (1) power consumption, (2) economical revenue, (3) resource utilization and (4) placement reconfiguration time.


For IaaS customers, cloud computing resources often appear to be unlimited and can be provisioned in any quantity at any required time \cite{Mell2011}. Consequently, this work considers a basic federated-cloud deployment architecture for the VMP problem, and follows the randomness behavior of customer requests by the use of a scenario-based uncertainty approach for modeling several uncertain parameters \cite{jortigoza2015}.

\section{Motivation}

Since current cloud computing markets represents dynamic environments where different parameters could change through cloud applications life-cycle, from an Infrastructure as a Service (IaaS) provider's perspective, the VMP problem must be formulated as an online problem (considering that requests from customers are unknown a priori) and it should be solved with short time constraints.

An online problem formulation is considered when an algorithm makes decisions on-the-fly without knowledge of upcoming events (e.g. online heuristics for VMP problems) \cite{beloglazov2012optimal}. On the other hand, if an algorithm has a complete knowledge of the future events of a problem instance, the formulation is considered as offline (e.g. MAs for VMP problems used in \cite{Ihara2015f} and \cite{lopez2015b}).

Although online decisions can be quickly achieved they can also negatively affect the quality of the solutions in VMP problems when comparing to offline decisions. Unfortunately, offline approaches cannot be used in dynamic environments of VMP problems since VM requests information are not known beforehand.

To help IaaS providers in the design and implementation of resource allocation algorithms considering many objective functions, the following research question must be answered: \textit{Which heuristics are most suitable for solving online VMP problems in federated-clouds considering many objectives?}

A provider-oriented VMP problem formulation is proposed for the optimization of four objective functions: (1) power consumption, (2) economical revenue, (3) quality of service and (4) resource utilization (see Chapter \ref{chap:problem_formulation_first_part}), in order to compare the considered algorithms (see Chapter \ref{chap:algorithms_first_part}). To combine the above mentioned objectives, a weighted sum method is considered. Experiments were performed to define appropriate weights for each objective function (see Chapter \ref{chap:experimental_results_first_part}).

\chapter{Provider-oriented VMP Problem Formulation}
\label{chap:problem_formulation_first_part}

This chapter proposes a formulation of a provider-oriented VMP problem considering the optimization of the following objective functions: (1) power consumption, (2) economical revenue, (3) quality of service and (4) resource utilization. According to the taxonomy presented in \cite{lopez2015}, this work focuses on a provider-oriented VMP for federated-cloud deployments, excluding elasticity and overbooking of resources, considering two formulation types: (1) online and (2) offline.

The formulation of the proposed provider-oriented VMP problem is based on \cite{Ihara2015f} and could be enunciated as:

\vspace{2mm}

\noindent \textit{Given a cloud infrastructure composed by a set of PMs $(H)$, a dynamic scenario composed by a set of VMs requested at each discrete time $t$ $(V(t))$ and the current placement of VMs into PMs $(P(t))$, it is incrementally sought a placement of $V(t)$ into $H$ for the next time instant $(P(t+1))$, satisfying the constraints and optimizing the considered objectives.}

In online algorithms for solving the proposed VMP problem, placement decisions are performed at each discrete time $t$, without knowledge of upcoming VM requests (Figure \ref{fig:simple_vmp_scheme}). On the other hand, an offline algorithm has knowledge of the complete set of VM requests in order to decide the placement of these VMs into available PMs. Consequently, the current placement is not necessary optimal for offline algorithms.

\section{Input Data}
\label{sec:inputData}

The proposed formulation of the VMP problem models a cloud computing datacenter, composed by PMs and VMs, receiving the following information as input data:

\begin{itemize}
	
	\item a set of $n$ available PMs and their specifications (see \eqref{eq:H});
	
	\vspace{-2mm}
	
	\item a set of $m(t)$ VMs at each discrete time $t$ and their specifications (see \eqref{eq:V_t});
	
	\vspace{-2mm}
	
	\item the current placement at each discrete time $t$ (see \eqref{eq:P_t}).
\end{itemize}

The set of PMs is represented as a matrix $H \in \mathbb{R}^{n\times4}$, as presented in (Equation \ref{eq:H}). Each PM $H_i$ is represented by its physical resources. This work considers 3 different physical resources ($Pr_1$-$Pr_3$): CPU [ECU], RAM memory [GB] and network capacity [Mbps]. It is important to mention that the proposed notation is general enough to include more characteristics associated to physical resources. Finally, the maximum power consumption [W] is also considered, i.e.

\vspace{-2mm}
\begin{equation}
\label{eq:H}
    H=
    \left[
    \begin{array}{cccc}
        Pr_{1,1} & Pr_{2,1} & Pr_{3,1} & pmax_1 \\
		\dots  & \dots  & \dots  & \dots  \\
		Pr_{1,n} & Pr_{2,n} & Pr_{3,n} & pmax_n \\
    \end{array}
    \right]
\end{equation}

\vspace{-2mm}
\noindent \textit{where:}

\vspace{-2mm}
\begin{tabbing}
\hspace{1.5cm}\=\kill
    $i$:            \>PM identifier ($1 \leq i \leq n$);\\
    $Pr_{1,i}$:     \>Processing resources of $H_i$ in [ECU];\\
    $Pr_{2,i}$:     \>Memory resources of $H_i$ in [GB];\\
    $Pr_{3,i}$:     \>Network capacity resources of $H_i$ in [Mbps];\\
    $p_{max_i}$:    \>Maximum power consumption of $H_i$ in [W];\\
    $n$:            \>Number of PMs.
\end{tabbing}

The set of VMs requested by customers at each discrete time $t$ is represented as a matrix $V(t) \in \mathbb{R}^{m\times5}$, as presented in (Equation \ref{eq:V_t}). In this work, each VM $V_{j}$ requires 3 different virtual resources ($Vr_1$-$Vr_3$): CPU [ECU], RAM memory [GB] and network capacity [Mbps]. It is important to mention that the notation could represent any other set of virtual resources such as: Block Storage or even Graphics Processing Unit (GPU). An economical revenue $R_j$ [USD] is associated to each VM $V_{j}$ as well as a priority level represented as a Service Level Agreement (SLA). The VMs try to lease the requested virtual resources for a fixed period of discrete time.

\vspace{-2mm}
\begin{equation}
\label{eq:V_t}
 	V(t)= \left[
		\begin{array}{ccccc}
			Vr_{1,1} & Vr_{2,1} & Vr_{3,1} & SLA_1 & R_1 \\
			\dots & \dots & \dots & \dots & \dots \\
      Vr_{{1,m(t)}} & Vr_{{2,m(t)}} & Vr_{{3,m(t)}} & SLA_{m(t)} & R_{m(t)}
    \end{array}
	\right]
\end{equation}

\vspace{-2mm}
\noindent \textit{where:}

\vspace{-2mm}
\begin{tabbing}
\hspace{1.5cm}\=\kill
    $j$:            \>VM identifier ($1 \leq j \leq m(t)$);\\
	$Vr_{1,j}$: 	\>Processing requirements of $V_j$ in [ECU]; \\
	$Vr_{2,j}$: 	\>Memory requirements of $V_j$ in [GB]; \\
	$Vr_{3,j}$:     \>Network requirements of $V_j$ in [Mbps]; \\
	$R_j$: 			\>Economical revenue for attending $V_j$ in [USD]; \\
	$SLA_j$: 		\>SLA of $V_j$, where $SLA_j \in \{1,2,\dots,s\}$, being
					$s$ the highest priority level; \\
	$m(t)$:			\>Number of VMs at each discrete time $t$, then $m(t) \in \{1,\dots,m_{max}\}$; \\
	$m_{max}$:      \>Maximum number of acceptable VMs.
\end{tabbing}

Once a VM $V_{j}$ is powered-off by the customer, its virtual resources are released, so the provider can reuse them. For simplicity, in what follows the index $j$ is not reused; therefore, for this work $V_{j}$ is not a function of time.

The current placement of VMs into PMs $(P(t))$ is also considered as an input data, taking into account that the placement of requested VMs is performed incrementally at each discrete time $t$. The placement at each discrete time $t$ is represented as a matrix $P(t) \in \mathbb{R}^{n \times m(t)}$, defined as:

\vspace{-2mm}
\begin{equation}
\label{eq:P_t}
	P(t)= \left[
	\begin{array}{cccc}
		P_{1,1}(t) & P_{1,2}(t) & \dots & P_{1,n}(t) \\
		\dots & \dots & \dots & \dots \\
		P_{m(t),1}(t) & P_{m(t),2}(t) & \dots & P_{m(t),n}(t)
	\end{array}
\right]
\end{equation}

\vspace{-2mm}
\noindent \textit{where:}

\vspace{-2mm}
\begin{tabbing}
\hspace*{3cm}\=\kill
    $P_{j,i}(t) \in \{0,1\}$:   \>Indicates if $V_j$ is allocated $(P_{j,i}(t) = 1)$ or not $(P_{j,i}(t) = 0)$\\
                                \>for execution on a PM $H_i$ (i.e., $P_{j,i}(t) : V_j \rightarrow H_i)$ at a discrete time $t$.
\end{tabbing}

\section{Output}
\label{sec:output}

The result of the proposed VMP problem at each discrete time $t$ is a new placement for the next time instant $(P(t+1))$. This is represented by a matrix $P(t+1) \in \mathbb{R}^{n \times m(t+1)}$.

\vspace{-2mm}

\begin{equation} \label{eq:P_t1}
    P(t+1)= \left[ 
    \begin{array}{ccc}
        P_{1,1}(t+1) & \dots & P_{1,n}(t+1) \\
        \dots & \dots & \dots \\
        P_{m(t+1),1}(t+1) & \dots & P_{m(t+1),n}(t+1) 
    \end{array}
    \right]
\end{equation}

\vspace{-2mm}
\noindent \textit{where:}

\vspace{-2mm}
\begin{tabbing}
\hspace*{3.5cm}\=\kill
    $P_{j,i}(t+1) \in \{0,1\}$: \>Indicates if $V_j$ is allocated $(P_{j,i}(t) = 1)$ or not $(P_{j,i}(t) = 0)$\\
                                \> for execution on a PM $H_i$ (i.e., $P_{j,i}(t) : V_j \rightarrow H_i)$
                                at instant $t+1$.
\end{tabbing}

\vspace{-2mm}

\begin{figure}[!ht]
	\centering
	\includegraphics[scale=0.37]{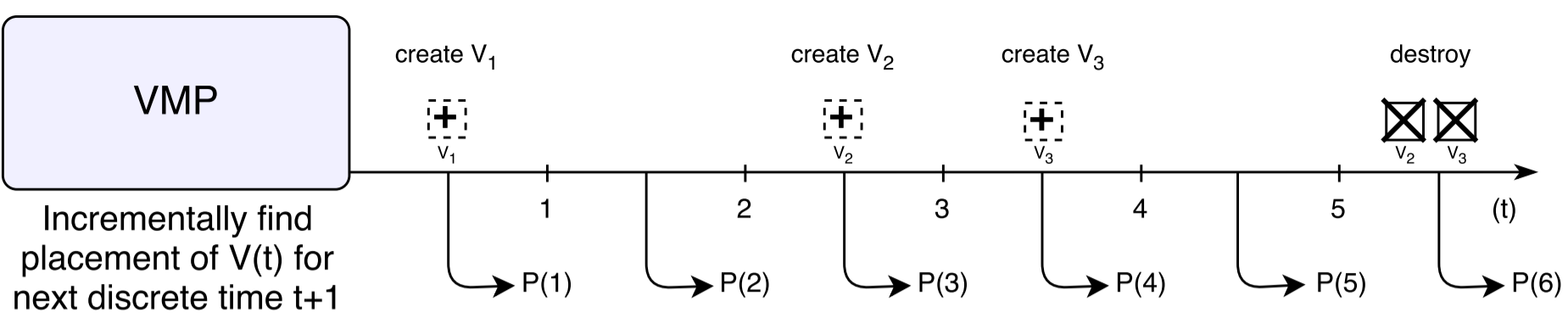}
	\caption{Allocation process for VMP problems considered in this work, presenting a basic example with VM requests arriving at different times (from $t=0$ to $t=4$) and VM resources being released (after $t=5$) and how those actions affect the placement $P$ as the times advances.}
	\label{fig:simple_vmp_scheme}
\end{figure}

Figure \ref{fig:simple_vmp_scheme} illustrates how VMs requests are allocated and released dynamically in cloud computing datacenters, 
where placement decisions are performed at each discrete time $t$, without knowledge of upcoming VM requests.

\section{Constraints}
\label{sec:constraints}

Several constraints should be considered when solving a VMP problem, they are described in this section.

\subsection{Constraint 1: Unique Placement of VMs}
\label{subsec:constraint_1}
A VM $V_{j}$ should be located to run on a single PM $H_i$ or alternatively for each $V_j$ such that $SLA_j < s$, it could not be located into any PM. Consequently, this placement constraint is expressed as:

\vspace{-2mm}
\begin{equation}
\label{eq:unique_placement}
	\begin{aligned}
		&\sum_{i=1}^{n} P_{j,i}(t) \le 1 \quad \forall j \in \{1,\dots,m(t)\}
	\end{aligned}
\end{equation}

\subsection{Constraint 2: Assure SLA Provisioning}
\label{subsec:constraint_sla}

A VM $V_{j}$ with the highest level of SLA (i.e. $SLA_j = s$) must be mandatorily allocated to run on a PM $H_i$. Consequently, this constraint is mathematically expressed as:

\vspace{-2mm}
\begin{equation}
\label{eq:sla_provisioning}
	\begin{aligned}
		&\sum_{i=1}^{n} P_{j,i}(t) = 1 \quad \forall j \text{ such that } SLA_j=s
	\end{aligned}
\end{equation}

\subsection{Constraints 3-5: Physical Resources of PMs}
\label{subsec:constraint3-5_part_one}

A PM $H_i$ must have sufficient available resources to meet the requirements of all VMs $V_{j}$ that are allocated to run on $H_i$. Consequently, these constraints can be formulated as:

\vspace{-2mm}
\begin{equation}
\label{eq:phys_cpu_first_part}
	\begin{aligned}
		\sum_{j=1}^{m(t)} Vr_{k,j} \times P_{j,i}(t) \le Pr_{k,i}
	\end{aligned}
\end{equation}

\vspace{-2mm}
\begin{center}
    $\forall i \in \{1,\dots,n\}$ and $\forall k \in \{1,\dots,r\}$, \\
    i.e. for each PM $H_i$ and for each considered resource $r$.
\end{center}

Physical resources are considered as resources available for VMs, without considering resources for the PM's hypervisor.

\section{Objective Functions}
\label{sec:objectiveFunctions_part_one}

Each of the considered objective functions must be formulated in a single optimization context (i.e. minimization or maximization) and each objective function's cost must be normalized to be comparable and combinable as a single objective. This work normalizes each objective function cost by calculating $\hat{f_i}(x) \in \mathbb{R}$, where $0 \leq \hat{f_i}(x) \leq 1$ (See Section \ref{sec:weighted_sum_method}).

\subsection{Power Consumption Minimization}
\label{subsec:powerConsumption}
Based on Beloglazov et al. \cite{beloglazov2012energy}, this work models the power consumption of PMs considering a linear relationship with the CPU utilization of PMs. This can be represented by the sum of the power consumption of each PM $H_i$.

\vspace{-2mm}

\begin{equation} 
\label{eq:min_power_consumption_part_one}
	min \hspace{1mm} f_1(x) = \sum_{i=1}^{n} ((p_{max_i} - p_{min_i}) \times Ur_{1,i}(t) + p_{min_i}) \times Y_i(t)
\end{equation}

\vspace{-2mm}
\noindent \textit{where:}

\vspace{-2mm}
\begin{tabbing}
\hspace*{3cm}\=\kill
    $x$:                    \>Evaluated solution of the problem $P(t)$; \\
	$f_1(x)$:               \>Total power consumption of PMs; \\
	$p_{min_i}$:            \>Minimum power consumption of a PM $H_i$.\\
                            \>As suggested in \cite{beloglazov2012energy}, $p_{min_i} \approx p_{max_i} \times 0.6$; \\
	$Ur_{1,i}(t)$:          \>Utilization ratio of resource 1 (in this case CPU) by $H_i$ at instant
	                        $t$;\\
	$Y_i(t) \in \{0,1\}$:   \>Indicates if $H_i$ is turned on $(Y_i(t) = 1)$ or not $(Y_i(t) = 0)$ at                            instant $t$.
\end{tabbing}

\subsection{Economical Revenue Maximization}
\label{subsec:economicalRevenue_part_1}

For IaaS customers, cloud computing resources often appear to be unlimited and can be provisioned in any quantity at any required time $t$ \cite{Mell2011}. Consequently, this work considered a federated-cloud deployment architecture, where a main provider may attend to requested resources that are not able to be provided (e.g. a workload peak) by transparently leasing low-price resources at a federated provider \cite{6779415}.

In this work, the maximization of the total economical revenue that a provider receives from attending the requirements of its customers is achieved by minimizing the total costs of leasing resources from alternative datacenters of the cloud federation. Equation (\ref{eq:economical_revenue_part_one}) represents the mentioned leasing costs, defined by the sum of the total costs of leasing each VM $V_j$ that is effectively allocated for execution on any PM of an alternative datacenter of the cloud federation.

A provider must offer its idle resources to the cloud federation at lower prices than offered to customers in the actual cloud market. The pricing scheme may depend on the particular agreement between providers of the cloud federation \cite{6779415}. Consequently, this work considers that the main provider may lease requested resources (that are not able to be provided) from the cloud federation at 70\% $(\hat{X_j} = 0.7)$ of its price in markets ($R_j$). This objective may be formulated as:

\vspace{-2mm}
\begin{equation}
\label{eq:economical_revenue_part_one}
	min \hspace{1mm} f_2(x) = \sum_{j=1}^{m(t)} (R_j \times X_j(t) \times \hat{X_j})
\end{equation}

\vspace{-2mm}
\noindent	\textit{where:}

\vspace{-2mm}
\begin{tabbing}
\hspace*{3cm}\=\kill
	$f_2(x)$:               \>Total costs for not allocating VMs in the main provider;\\
	$X_j(t)\in \{0,1\}$:    \>Indicates if $V_j$ is allocated for execution on a PM $(X_j(t) = 1)$\\
	                        \>or not $(X_j(t) = 0)$ at instant $t$; \\
	$\hat{X_j}$:            \>Indicates if $V_j$ is allocated on the main provider $(\hat{X_j} = 0)$\\
	                        \>or on an alternative datacenter of the cloud federation $(\hat{X_j} = 0.7)$.
\end{tabbing}

It is important to note that $\hat{X_j}$ is not a function of time. The decision of locating a VM $V_j$ on a federated provider is considered only in the placement process, with no possible migrations between providers. The value of $\hat{X_j}$ depends on the agreement celebrated between federated providers.

\subsection{Quality of Service Maximization}
\label{subsec:qos}

In this work, the quality of service (QoS) maximization proposes the allocation of the maximum number of VMs with the highest level of priority associated to the SLA prior to the VMs with smaller SLA priority. In case the main provider allocates a VM in a federated provider the QoS is considered as 0. In order to evaluate this objective in a minimization context, the total of SLA violations is minimized and formulated as:

\vspace{-2mm}
\begin{equation}
\label{eq:qos_part_one}
    min \hspace{1mm} f_3(x) = \sum_{j=1}^{m(t)} (\hat{C}^{SLA_j} \times SLA_j \times X_j(t) \times \hat{Y_j})
\end{equation}

\vspace{-2mm}
\noindent \textit{where:}

\vspace{-2mm}
\begin{tabbing}
\hspace*{2.4cm}\=\kill
    $f_3(x)$:               \>Total SLA violations figure for a given placement; \\
    $\hat{C}$:              \>Constant, large enough to prioritize services with a larger $SLA$\\
                            \> over the ones with a lower $SLA$; \\
	$\hat{Y_j}\in \{0,1\}$: \>Indicates if $V_j$ is allocated for its execution on the main provider $(\hat{Y_j} = 0)$\\
                            \>or on an alternative datacenter of the cloud federation $(\hat{Y_j} = 1)$.
\end{tabbing}

\subsection{Resources Utilization Maximization}
\label{subsec:res_wasted}
An efficient utilization of resources is a relevant management challenge to be addressed by IaaS providers. This work proposes the resource utilization maximization strategy by minimizing the average ratio of wasted resources on each PM $H_i$ (i.e. resources that are not allocated to any VM $V_j$). This objective function is formulated in Equation (\ref{eq:res_wasted_min}).

\vspace{-2mm}

\begin{equation}
\label{eq:res_wasted_min}
    min \hspace{1mm} f_4(x) = 
		\frac{\sum_{i=1}^{n}
		\bigg[1-
			\bigg(
				\dfrac{Ur_{1,i}(t) + \dots + Ur_{r,i}(t)}{r}
			\bigg)
		\bigg] 
		\times Y_i(t)}
		{\sum_{i=1}^{n} Y_i(t)}
\end{equation}

\vspace{-2mm}
\noindent \textit{where:}

\vspace{-2mm}
\begin{tabbing}
\hspace*{3cm}\=\kill
    $f_4(x)$:       \>Average ratio of wasted resources; \\
	$Ur_{1,i}(t)$:  \>Utilization ratio of resource 1 (in this case CPU) \\
                    \>by $H_i$ at instant $t$; \\
	$Ur_{r,i}(t)$:  \>Utilization ratio of resource $r$ (any resource) by \\
                    \>$H_i$ at instant $t$; \\
	$r$:            \>Number of considered resources. In this paper 3: \\
                    \>CPU, RAM memory and network capacity;\\
\end{tabbing}

The following section summarizes the main considerations taken into account to combine the four presented objective functions into a single objective.

\section{Normalization and Scalarization Method}
\label{sec:weighted_sum_method}

Experimental results presented in \cite{Ihara2015f} recommend the combination of many normalized objective functions into a single objective (e.g. minimum distance to origin) for IaaS environments. The first part of this work considers a WS method to combine many objectives into a single objective.

As previously mentioned, each objective function cost must is normalized to be comparable and combinable as a single objective. This work normalizes each objective function cost by calculating $\hat{f_i}(x) \in \mathbb{R}$, where $0 \leq \hat{f_i}(x) \leq 1$, as defined in \cite{salem2006high}:

\vspace{-2mm}
\begin{equation}
\label{eq:norm_FO}
	\hat{f_i}(x) = \frac{f_i(x) - f_i(x)_{min}}{f_i(x)_{max} - f_i(x)_{min}}
\end{equation}

\vspace{-2mm}
\noindent \textit{where:}

\vspace{-2mm}
\begin{tabbing}
\hspace*{3cm}\=\kill
	$\hat{f_i}(x)$: \>Normalized cost of objective function $f_i(x)$; \\
	${f_i}(x)$:     \>Cost of objective function $f_i(x)$; \\
	$f_i(x)_{min}$:	\>Minimum possible cost for $f_i(x)$; \\
	$f_i(x)_{max}$:	\>Maximum possible cost for $f_i(x)$. \\
	
\end{tabbing}

Finally, the four previously presented normalized objective functions are combined into a single objective considering a weighted sum method, expressed as follows:

\vspace{-2mm}
\begin{equation}
\label{eq:uof}
	\begin{aligned}
		min \hspace{1mm} f(x) = \sum_{i=1}^{q} \hat{f_i}(x) \times w_i
	\end{aligned}
\end{equation}

\vspace{-2mm}
\noindent \textit{where:}

\vspace{-2mm}
\begin{tabbing}
\hspace{3cm}\=\kill
	${f}(x)$:       \>Single objective function combining each $f_i(x)$; \\
	$w_i$:          \>Weight of importance associated to $f_i(x)$; \\
	$q$:            \>Number of objective functions. In this case 4.
\end{tabbing}

In this work, values of each weight $w_i$ associated to an objective function $f_i(x)$ were obtained empirically by analyzing a large number of experiments to be presented in Chapter \ref{chap:experimental_results_first_part}. 

\chapter{Evaluated Algorithms}
\label{chap:algorithms_first_part}

To analyze alternatives to solve the formulated VMP problem (see Chapter \ref{chap:problem_formulation_first_part}), an experimental comparison of five different online deterministic heuristics against an offline memetic algorithm with migration of VMs was performed. This chapter summarizes the five online deterministic heuristic algorithms proposed for the experimental comparison against the presented offline memetic algorithm.

\section{A1: First-Fit (FF)}
\label{sec:firstFit}
In the FF algorithm, requested VMs $V_j$ are allocated on the first PM $H_i$ with available resources (see Section \ref{sec:constraints}). Interested readers can refer to \cite{fang2013power} for details on FF, BF and WF algorithms applied to VMP problems. (See Algorithm \ref{alg_ff}).

\begin{algorithm}[!ht]
	\SetKwFunction{Union}{Union}
	\SetKwFunction{UpdatePlacementMatrix}{UpdatePlacementMatrix}

	\SetAlgoLined
	\small
	\KwData{Datacenter infrastructure (see Section \ref{sec:inputData})}
	\KwResult{true, if the VM is allocated correctly. Otherwise false}
	\ForEach{$H_i$ where $i \le n$}{
		\If{$H_i$ has enough resources to host VM}{
			allocate VM into PM\;
			\UpdatePlacementMatrix{$H_i$}\;
			return true\;
		}
	}
	return false\;
 	\caption{First-Fit Algorithm}
 	\label{alg_ff}
\end{algorithm}

\section{A2: Best-Fit (BF)}
\label{sec:bestFit}
The BF algorithm, allocates requested VMs $V_j$ on the first PM $H_i$ with available capacity from a sorted list of PMs in increasing order by a score associated to each PM. The score of a PM $H_i$ is determined as a sum of its ratios of unutilization of resources, as detailed in \cite{fang2013power}. (See Algorithm \ref{alg_bf}).

\begin{algorithm}[!ht]
	\SetKwFunction{Union}{Union}\SetKwFunction{UpdatePlacementMatrix}{UpdatePlacementMatrix}
	\SetKwFunction{Union}{Union}\SetKwFunction{CalculatePMWeight}{CalculatePMWeight}
	\SetKwFunction{Union}{Union}\SetKwFunction{UpdatBestFitSortedList}{UpdatBestFitSortedList}

	\SetAlgoLined
	\small
	\KwData{Datacenter infrastructure (see Section \ref{sec:inputData})}
	\KwResult{true, if the VM is allocated correctly. Otherwise, false.}
	\ForEach{$H_i$ where $i \le n$}{
		weight = \CalculatePMWeight{$H_i$}\;
		\UpdatBestFitSortedList{$H_i$, weight}\;
	}
	\ForEach{$H_i$ in $sortedList$}{
		\If{$H_i$ has enough resources to host $V_j$}{
			allocate $V_j$ into $H_i$\;
			\UpdatePlacementMatrix{$H_i$}\;
			return true\;
		}
	}
	return false\;
	\caption{Best-Fit Algorithm}
	\label{alg_bf}
\end{algorithm}

\section{A3: Worst-Fit (WF)}
\label{sec:worstFit}

In the WF algorithm, VMs $V_j$ are allocated on the first available PM $H_i$ of a decreasingly ordered list of PMs based on the score of the PM, inversely to the operation of the BF algorithm. For more details, refer to \cite{fang2013power}. (See Algorithm \ref{alg_wf}).

\begin{algorithm}[!ht]
	\SetKwFunction{Union}{Union}\SetKwFunction{UpdatePlacementMatrix}{UpdatePlacementMatrix}
	\SetKwFunction{Union}{Union}\SetKwFunction{CalculatePMWeight}{CalculatePMWeight}
	\SetKwFunction{Union}{Union}\SetKwFunction{UpdateWorstFitSortedList}{UpdateWorstFitSortedList}

	\SetAlgoLined
	\small
	\KwData{Datacenter infrastructure (see Section \ref{sec:inputData})}
	\KwResult{true, if the VM is allocated correctly. Otherwise, false.}
	\ForEach{$H_i$ where $i \le n$}{
		weight = \CalculatePMWeight{$H_i$}\;
		\UpdateWorstFitSortedList{$H_i$, weight}\;
	}
	\ForEach{$H_i$ is $sortedList$}{
		\If{$H_i$ has enough resources to host $V_j$}{
			allocate $V_j$ into $H_i$\;
			\UpdatePlacementMatrix{$H_i$}\;
			return true\;
		}
	}
	return false\;
	\caption{Worst-Fit Algorithm}
	\label{alg_wf}
\end{algorithm}

\section{A4: First-Fit Decreasing (FFD)}
\label{sec:firstFitDecreasing}

The First-Fit Decreasing (FFD) algorithm operates similarly to the previously presented FF algorithm. The main difference with the FF algorithm is that the FFD algorithm sorts the list of requested VMs $V_j$ in decreasing order by requested CPU resources, as described in \cite{Ferreto2011}. (See Algorithm \ref{alg_ffd}).

\begin{algorithm}[!ht]
	\SetKwFunction{Union}{Union}
	\SetKwFunction{sortRequestsInDecreasingOrder}{sortReqInDescOrder}
	\SetKwFunction{firstFit}{firstFit}

	\SetAlgoLined
	\small
	\KwData{Datacenter infrastructure (see Section \ref{sec:inputData})}
	\KwResult{true, if the VM is allocated correctly. Otherwise, false.}

    sortedRequestList = \sortRequestsInDecreasingOrder(requestList)\;
    \ForEach{request in sortedRequestList}{
    	\firstFit(request, $H$, $n$)\;
    }
	\caption{First-Fit Decreasing Algorithm}
	\label{alg_ffd}
\end{algorithm}

\section{A5: Best-Fit Decreasing (BFD)}
\label{sec:bestFitDecreasing}

The BFD algorithm has a similar behavior to the BF algorithm. The main difference between the BF algorithm and the BFD one is that the BFD algorithm sorts the list of requested VMs $V_j$ in decreasing order by requested CPU resources. Interested readers can refer to \cite{Dong2013} for details on the BFD algorithm applied to VMP problems. (See Algorithm \ref{alg_wfd}).

\begin{algorithm}[!ht]
\SetKwFunction{Union}{Union}
	\SetKwFunction{sortRequestsInDecreasingOrder}{sortReqInDescOrder}
	\SetKwFunction{bestFit}{bestFit}

	\SetAlgoLined
	\small
	\KwData{Datacenter infrastructure (see Section \ref{sec:inputData})}
	\KwResult{true, if the VM is allocated correctly. Otherwise, false.}

    sortedRequestList = \sortRequestsInDecreasingOrder(requestList)\;
    \ForEach{request in sortedRequestList}{
    	\bestFit(request, $H$, $n$)\;
    }
    
	\caption{Best-Fit Decreasing Algorithm}
	\label{alg_wfd}
\end{algorithm}

\section{A6: Memetic Algorithm}
\label{sec:offlineResolutionApproach_first_part}

A MA is considered for an offline alternative to solve the formulated VMP problem taking into account many objectives (see Chapter \ref{chap:problem_formulation_first_part}). This algorithm is based on the one proposed in \cite{Ihara2015f} by Ihara et al.

The considered algorithm may be classified as a meta-heuristic, following an evolutionary process to find appropriate solutions. Basically, the considered MA follows this evolutionary behavior: solutions are selected from an evolutionary set of solutions (or population). Crossover and mutation operators are applied as usual, and eventually solutions are repaired, as there may be unfeasible solutions. Improvements on the evolutionary population's solution may be generated using local optimization operators. Next, a new evolutionary population is selected from the union of the (until that generation) best population and the set of improved solutions. The evolutionary process is repeated until the algorithm meets a stopping criterion (such as a maximum number of generations), returning from the evolutionary population the solution with minimum cost of $f(x)$.  Details are presented in \cite{lopez2016LANC}.


\chapter{Part I - Experimental Results}
\label{chap:experimental_results_first_part}

In this work, an experimental comparison of five of the most studied online deterministic heuristics was performed considering several experimental workloads. The quality of solutions obtained by the evaluated online heuristics was compared to the average performance (in ten runs) of an offline MA that has complete knowledge of future VM requests and the ability to migrate VMs between PMs.

\section{Experimental Environment}
\label{sec:environment_first_part}

The six evaluated algorithms were implemented using ANSI C programming language. The source code is available online\footnote{\small http:$\slash$$\slash$github.com$\slash$DynamicVMP$\slash$vmpCompetitiveAnalysis}, as well as all the considered experimental data. Experiments were performed on a Linux Operating System with an Intel Core i7 2.3 GHz CPU and 12 GB of RAM memory.

Physical resources (matrix $H$) represent an homogeneous IaaS cloud composed by 10 PMs with the following specifications: 8 [ECU] of CPU, 10 [GB] of RAM memory, 780 [Mbps] of network capacity and 960 [W] of maximum power consumption (see Section \ref{sec:inputData} for notation details).

In this work, requested virtual resources (matrix $V(t)$) were considered using four different workload traces generated using a \textit{Workload Trace Generator} for provider-oriented VMP problems \cite{ortigoza2016} available for research purposes\footnote{\small http:$\slash$$\slash$wtg.cba.com.py\label{foot:wtg}}.

Each considered workload trace simulates customers requests to allocate a set of 100 VMs following different Probability Distribution Functions (PDFs) for the VM request arrivals. Three workload traces follow a Poisson PDF with different expected values ($\lambda$): ($W_1$): $\lambda = 10$, ($W_2$): $\lambda = 50$ and ($W_3$): $\lambda = 70$. In this case, different values of $\lambda$ may represent workload peaks at different time instants $t$ (see Figures \ref{fig:environment_W1} to \ref{fig:environment_W3}).

\begin{figure*}[!ht]
	\begin{tikzpicture}
		\begin{axis}[
				title={},
				height=20em,
				width=39em,
				tickpos=left,
				ymin=0,
				ymax=16,
				xmin=0,
				xmax=110,
				bar width=0.02cm,
				tick label style={font=\scriptsize},
				enlarge x limits=0.00,
				xtick={0,10,20,30,40,50,60,70,80,90,100},
				ytick={0,3,6,9,12,15},
				xlabel={Time ($t$)},
				ylabel={Number of created VMs},
				legend style={legend pos=north east,font=\tiny},
			]
			\addlegendentry[text width=65pt, text depth=]{$W_1$: Poisson $\lambda = 10$}
			
			\addplot[mark=*,only marks,red] coordinates {
				(0,0)(1,1)(2,0)(3,3)(4,2)(5,6)(6,9)(7,6)(8,10)(9,14)(10,11)(11,6)(12,8)(13,7)
				(14,6)(15,6)(16,4)(17,1)(18,0)(99,0)
			};
		\end{axis}
	\end{tikzpicture}
	
	\caption{Experimental Workload Trace: Number of created VMs as a function of time $t$ utilizing a Poisson distribution function with a parameter $\lambda = 10$.}
	\label{fig:environment_W1}
\end{figure*}

\begin{figure*}[!ht]
	\begin{tikzpicture}
		\begin{axis}[
				title={},
				height=20em,
				width=39em,
				tickpos=left,
				ymin=0,
				ymax=16,
				xmin=0,
				xmax=110,
				bar width=0.02cm,
				tick label style={font=\scriptsize},
				enlarge x limits=0.00,
				xtick={0,10,20,30,40,50,60,70,80,90,100},
				ytick={0,3,6,9,12,15},
				xlabel={Time ($t$)},
				ylabel={Number of created VMs},
				legend style={legend pos=north east,font=\tiny},
			]
		
			\addlegendentry[text width=65pt, text depth=]{$W_2$: Poisson $\lambda = 50 $}
			
			\addplot[mark=diamond*,only marks,blue] coordinates {
				(0,0)(29,0)(30,1)(31,0)(32,0)(33,1)(34,0)(35,0)(36,1)(37,2)(38,2)
				(39,0)(41,0)(42,5)(43,3)(44,4)(45,4)(46,10)(47,4)(48,3)(49,5)(50,7)(51,5)(52,7)(53,9)(54,3)
				(55,4)(56,7)(57,5)(58,1)(60,1)(61,2)(62,1)(63,1)(64,1)(65,0)(99,0)
			};
		\end{axis}
	\end{tikzpicture}
	
	\caption{Experimental Workload Trace: Number of created VMs as a function of time $t$ utilizing a Poisson distribution function with a parameter $\lambda = 50$.}
	\label{fig:environment_W2}
\end{figure*}

\begin{figure*}[!ht]
	\begin{tikzpicture}
		\begin{axis}[
				title={},
				height=20em,
				width=39em,
				tickpos=left,
				ymin=0,
				ymax=16,
				xmin=0,
				xmax=110,
				bar width=0.02cm,
				tick label style={font=\scriptsize},
				enlarge x limits=0.00,
				xtick={0,10,20,30,40,50,60,70,80,90,100},
				ytick={0,3,6,9,12,15},
				xlabel={Time ($t$)},
				ylabel={Number of created VMs},
				legend style={legend pos=north east,font=\tiny},
			]
		
			\addlegendentry[text width=65pt, text depth=]{$W_3$: Poisson $\lambda = 70$}
			
			\addplot[mark=square*,only marks,teal] coordinates {
				(0,0)(47,0)(48,1)(49,1)(50,0)(51,0)(52,1)(53,0)(54,0)(55,1)(56,2)(57,2)(58,2)(59,4)(60,1)
				(61,2)(62,4)(63,3)(64,2)(65,3)(66,3)(67,5)(68,11)(69,7)(70,2)(71,5)(72,4)(73,2)(74,5)(75,5)
				(76,4)(77,4)(78,2)(79,1)(80,4)(81,2)(82,3)(83,0)(88,0)(89,1)(90,1)(91,0)(99,0)
			};
		\end{axis}
	\end{tikzpicture}
	
	\caption{Experimental Workload Trace: Number of created VMs as a function of time $t$ utilizing a Poisson distribution function with a parameter $\lambda = 70$.}
	\label{fig:environment_W3}
\end{figure*}

The fourth workload trace ($W_4$) follows an Uniform PDF, representing a stable workload of VM requests. All parameters considered for generating the experimental workload traces are presented in Table \ref{tab:inputs}. Interested readers can refer to \cite{ortigoza2016} for more details.

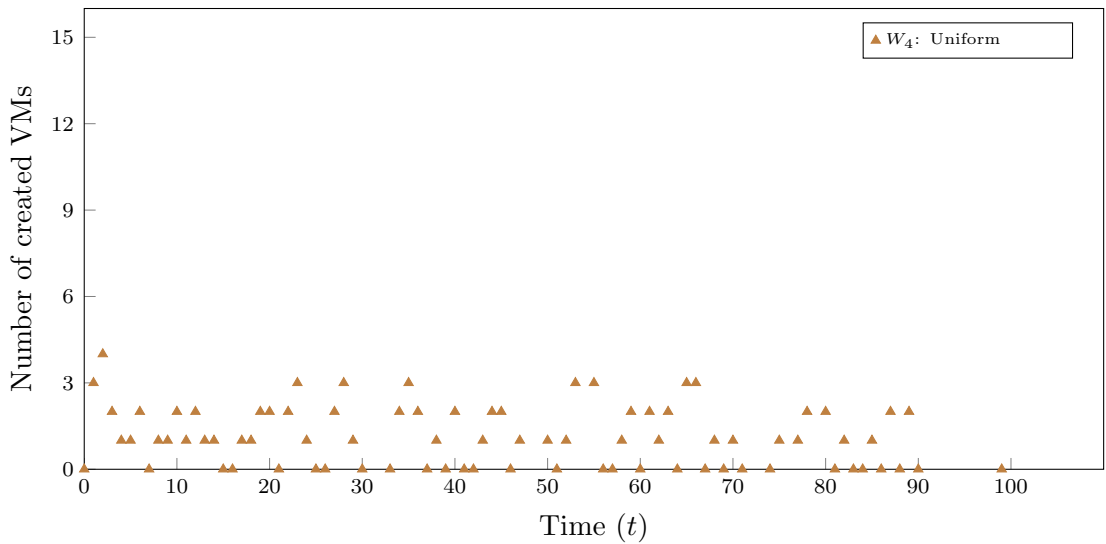
\begin{figure*}[!ht]
	\begin{tikzpicture}
		\begin{axis}[
				title={},
				height=20em,
				width=39em,
				tickpos=left,
				ymin=0,
				ymax=16,
				xmin=0,
				xmax=110,
				bar width=0.02cm,
				tick label style={font=\scriptsize},
				enlarge x limits=0.00,
				xtick={0,10,20,30,40,50,60,70,80,90,100},
				ytick={0,3,6,9,12,15},
				xlabel={Time ($t$)},
				ylabel={Number of created VMs},
				legend style={legend pos=north east,font=\tiny},
			]
		
			\addlegendentry[text width=65pt, text depth=]{$W_4$: Uniform}
			
			\addplot[mark=triangle*,only marks,brown] coordinates {
				(0,0)(1,3)(2,4)(3,2)(4,1)(5,1)(6,2)(7,0)(8,1)(9,1)(10,2)
				(11,1)(12,2)(13,1)(14,1)(15,0)(16,0)(17,1)(18,1)(19,2)(20,2)(21,0)(22,2)(23,3)(24,1)
				(25,0)(26,0)(27,2)(28,3)(29,1)(30,0)(33,0)(34,2)(35,3)(36,2)(37,0)(38,1)(39,0)
				(40,2)(41,0)(42,0)(43,1)(44,2)(45,2)(46,0)(47,1)(50,1)(51,0)(52,1)(53,3)(55,3)
				(56,0)(57,0)(58,1)(59,2)(60,0)(61,2)(62,1)(63,2)(64,0)(65,3)(66,3)(67,0)(68,1)
				(69,0)(70,1)(71,0)(74,0)(75,1)(77,1)(78,2)(80,2)(81,0)(82,1)(83,0)(84,0)(85,1)(86,0)(87,2)
				(88,0)(89,2)(90,0)(99,0)
			};
		\end{axis}
	\end{tikzpicture}
	
	\caption{Experimental Workload Trace: Number of created VMs as a function of time $t$ utilizing a Uniform distribution function}
	\label{fig:environment_W4}
\end{figure*}

\begin{table}[!ht]
\caption{Considered Data to generate Experimental Workload Traces ($W_1$ to $W_4$).}
\label{tab:inputs}

    \ra{1.3}
    \centering
	
	\begin{tabular}{@{}ll@{}}
	\toprule
	\bfseries Parameter Description & \bfseries Input Data \\
	\midrule
	1. Environment & No overbooking, No elasticity \\ 

	2. Workload Trace Duration [$t$] & 100 \\ 

	3. Number of Parameter Datacenters & 1 \\

	4. Number of Cloud Services & 100 \\

	5. Number of VMs per Service & 1 \\
	
	\\

	\multirow{4}{*}{6. VMs creation time [$t$]} & $W_1$: Poisson($\lambda = 10$) \\
											    & $W_2$: Poisson($\lambda = 50$) \\
											    & $W_3$: Poisson($\lambda = 70$) \\
											    & $W_4$: Uniform(0,100) \\
											    
	\\

	7. CPU Resources [ECU] 	& Uniform(1,8) \\

	8. RAM Resources [GB] 	& Uniform(1,8) \\

	9. Network Capacity [Mbps] & Uniform(10,1000) \\

	10. Revenue of VMs ($R_j$) [\$] & Uniform(0.1,1.5) \\

	11. SLA Level of VMs ($SLA_j$) & Uniform(1,5) \\
	\bottomrule
	\end{tabular}
\end{table}

In order to effectively analyze the described federated-cloud deployment architecture and provider-oriented VMP formulation considering many objectives, experimental workload traces requested more virtual resources than the available ones in the considered main cloud computing datacenter.

Experiments could be summarized as: For each considered workload trace ($W_1$ to $W_4$), one run of the following deterministic heuristics was performed in an online context: (1) FF, (2) BF, (3) WF, (4) FFD and (5) BFD. Considering the randomness of the MA compared against the previously mentioned heuristics, ten runs of the mentioned algorithm were performed. The average values of the ten runs were considered for the experimental comparison, as summarized in Table \ref{tab:WSvaluesPerAlgorithm}, where the offline MA clearly outperforms all five online heuristics as expected, given that it uses complete knownledge of VM requests and migration of VMs to optimize the objective function presented in Chapter \ref{chap:problem_formulation_first_part}.

\section{Objective Functions Weights}

To determine appropriate values for the weights $w_i$ associated to each objective function $f_i(x)$ (see Equation \eqref{eq:uof}), an exploration of the VMP problem domain was performed. In this context, 1000 feasible solutions ($x_1$ to $x_{1000}$) were randomly generated by the MA (A6) considering: $H$ as described in Section \ref{sec:environment_first_part}, $V(t)$ as presented in \textit{entorno00-1} (a \textit{benchmark} available online\footref{foot:wtg}), highest priority of VMs $s$ = 4 and $\hat{C}$ = 1000.

Obtained values of each objective function $f_i(x)$ were normalized in $\hat{f_i}(x)$, as described in Section \ref{sec:weighted_sum_method}. Consequently, each weight $w_i$ was defined as:

\vspace{-2mm}
\begin{equation}
\label{eq:w_i}
    w_i = \frac{1000}{\sum_{k=1}^{1000}\hat{f_i}(x_k)}
\end{equation}

\vspace{-2mm}
\noindent resulting in:
	$w_1$: \hspace{0.3mm} 1.3903; \quad $w_2$: \hspace{0.3mm} 2.1379; \quad $w_3$: \hspace{0.3mm} 2.7393; \quad $w_4$: \hspace{0.3mm} 1.4586.

\begin{table*}[!b]
	\caption{Objective Function Costs of Evaluated Heuristics}
	\label{tab:WSvaluesPerAlgorithm}
	
	\ra{1.3}
	\centering
	
	\resizebox{37em}{!} {
		\begin{tabular}{rcrrrrrr}
			\toprule
			& \phantom{abc} & \multicolumn{6}{c}{\textbf{Algorithms}}\\
			\cmidrule{3-8}
			\textbf{Workloads} && \textbf{A1: FF} & \textbf{A2: BF} & \textbf{A3: WF} & \textbf{A4: FFD} & \textbf{A5: BFD} & \textbf{A6: MA}\\
			\textit{$W_1$: Poisson} $\lambda = 10$ && 3.2927 & 3.3098 & 3.5250 & 3.0205 & 3.1392 & \cellcolor[HTML]{d3d3d3}\textbf{2.6096} \\
			\textit{$W_2$: Poisson} $\lambda = 50$ && 2.4602 & 2.5112 & 2.4811 & 2.4602 & 2.4555 & \cellcolor[HTML]{d3d3d3}\textbf{2.0001} \\
			\textit{$W_3$: Poisson} $\lambda = 70$ && 1.7054 & 1.6458 & 1.7054 & 1.6458 & 1.6458 & \cellcolor[HTML]{d3d3d3}\textbf{1.3588} \\
			\textit{$W_4$: Uniform} && 3.1875 & 3.1556 & 3.0489 & 3.0907 & 3.1556 &  \cellcolor[HTML]{d3d3d3}\textbf{2.3420} \\
			\\
			\textbf{Average} && 2.6615 & 2.6556 & 2.6901 & 2.5543 & 2.5990 & \cellcolor[HTML]{d3d3d3}\textbf{2.0776} \\
			\textbf{Ranking} && 5th & 4th & 6th & 2nd & 3th & \cellcolor[HTML]{d3d3d3}\textbf{1st}\\
			\bottomrule
        \end{tabular}
    }
\end{table*}

\section{Comparison of Online Heuristics}
\label{sec:heuristicsAnalysis}

This section summarizes the main findings obtained in the experimental comparison of algorithms for the VMP formulation with many objectives presented in Chapter \ref{chap:problem_formulation_first_part}. 

The main goal of the experimental comparison presented in this section is to define \textit{which heuristics are most suitable for solving online formulations of provider-oriented VMPs in federated-clouds considering many objective funtions}. 

Table \ref{tab:WSvaluesPerAlgorithm} summarizes the costs of the objective function $f(x)$, that combines the four considered objective functions (see Section \ref{sec:weighted_sum_method}). It is worth noting that according to the experimental results presented in Table \ref{tab:WSvaluesPerAlgorithm}, there is no evaluated heuristic that outperforms all other alternatives in all experimental workload traces. Consequently, Table \ref{tab:WSvaluesPerAlgorithm} also presents the average cost of objective function $f(x)$ as well as the corresponding ranking. Apparently, the FFD algorithm is the best heuristic among the five considered ones, followed closely by BFD (Accordingly, the MA is first in this ranking).

\begin{small}
	\begin{table*}[!ht]
		\caption{Objective Function Costs of Evaluated Algorithms}
		\label{tab:OFvaluesPerAlgorithm}
		
		\ra{1.3}
		\centering
		
		\resizebox{37em}{!} {
            \begin{tabular}{@{} *{7}{c}r @{}}
        		\toprule
        		& \multicolumn{1}{c}{} & \multicolumn{5}{c}{\textbf{Heuristics}} \\
        		\cmidrule(lr){3-7}
        		\textbf{Workload} & \textbf{$f_i(x)$} & \textbf{A1: FF} & \textbf{A2: BF} & \textbf{A3: WF} & \textbf{A4: FFD} & \textbf{A5: BFD} & \textbf{Best Heuristic} \\
        		\hline
        		\multirow{4}{*}{\rotatebox[origin=c]{90}{\textit{$W_1$}}}
        		& $f_1(x)$ & 0.9240 & 0.9107 & \cellcolor[HTML]{d3d3d3}\textbf{0.8929} & 0.9255 & 0.9208 & A3\\
        		& $f_2(x)$ & 0.0372 & 0.0373 & 0.0391 & \cellcolor[HTML]{d3d3d3}\textbf{0.0359} & \cellcolor[HTML]{d3d3d3}\textbf{0.0359} & A4,A5\\
        		& $f_3(x)$ & 0.6104 & 0.6139 & 0.6670 & \cellcolor[HTML]{d3d3d3}\textbf{0.5101} & 0.5611 & A4\\
        		& $f_4(x)$ & 0.1756 & 0.1933 & 0.2556 & 0.1778 & \cellcolor[HTML]{d3d3d3}\textbf{0.1679} & A5\\
        		\\
        		\multirow{4}{*}{\rotatebox[origin=c]{90}{\textit{$W_2$}}}
        		& $f_1(x)$ & 0.5900 & 0.5756 & \cellcolor[HTML]{d3d3d3}\textbf{0.5612} & 0.5900 & 0.5903 & A3\\
        		& $f_2(x)$ & \cellcolor[HTML]{d3d3d3}\textbf{0.0404} & 0.0406 & 0.0420 & \cellcolor[HTML]{d3d3d3}\textbf{0.0404} & \cellcolor[HTML]{d3d3d3}\textbf{0.0404} & A4,A5\\
        		& $f_3(x)$ & 0.4790 & 0.4975 & \cellcolor[HTML]{d3d3d3}\textbf{0.4715} & 0.4790 & 0.4790 & A3\\
        		& $f_4(x)$ & 0.1652 & 0.1789 & 0.2189 & 0.1652 & \cellcolor[HTML]{d3d3d3}\textbf{0.1618} & A5\\
        		\\
        		\multirow{4}{*}{\rotatebox[origin=c]{90}{\textit{$W_3$}}}
        		& $f_1(x)$ & \cellcolor[HTML]{d3d3d3}\textbf{0.4146} & 0.4198 & 0.4203 & 0.4198 & 0.4198 & A1\\
        		& $f_2(x)$ & 0.0243 & \cellcolor[HTML]{d3d3d3}\textbf{0.0235} & 0.0245 & \cellcolor[HTML]{d3d3d3}\textbf{0.0235} & \cellcolor[HTML]{d3d3d3}\textbf{0.0235} & A2,A4,A5\\
        		& $f_3(x)$ & 0.3155 & \cellcolor[HTML]{d3d3d3}\textbf{0.3003} & 0.3103 & \cellcolor[HTML]{d3d3d3}\textbf{0.3003} & \cellcolor[HTML]{d3d3d3}\textbf{0.3003} & A2,A4,A5\\
        		& $f_4(x)$ & 0.1457 & \cellcolor[HTML]{d3d3d3}\textbf{0.1296} & 0.1497 & \cellcolor[HTML]{d3d3d3}\textbf{0.1296} & \cellcolor[HTML]{d3d3d3}\textbf{0.1296} & A2,A4,A5\\
        		\\
        		\multirow{4}{*}{\rotatebox[origin=c]{90}{\textit{$W_4$}}}
        		& $f_1(x)$ & 0.8549 & \cellcolor[HTML]{d3d3d3}\textbf{0.8544} & 0.8694 & 0.8598 & \cellcolor[HTML]{d3d3d3}\textbf{0.8544} & A2,A5\\
        		& $f_2(x)$ & 0.0375 & 0.0364 & \cellcolor[HTML]{d3d3d3}\textbf{0.0356} & 0.0371 & 0.0364 & A3\\
        		& $f_3(x)$ & 0.5311 & 0.5283 & \cellcolor[HTML]{d3d3d3}\textbf{0.4884} & 0.4931 & 0.5283 & A3\\
        		& $f_4(x)$ & 0.3177 & 0.3034 & \cellcolor[HTML]{d3d3d3}\textbf{0.2919} & 0.3188 & 0.3034 & A3\\
        		\midrule
        		\multirow{4}{*}{\rotatebox[origin=c]{90}{\textbf{Average}}}
        		& $f_1(x)$ & 0.6959 & 0.6901 & \cellcolor[HTML]{d3d3d3}\textbf{0.6859} & 0.6988 & 0.6963 & A3\\
        		& $f_2(x)$ & 0.0349 & 0.0345 & 0.0353 & 0.0343 & \cellcolor[HTML]{d3d3d3}\textbf{0.0341} & A5\\
        		& $f_3(x)$ & 0.4841 & 0.4851 & 0.4844 & \cellcolor[HTML]{d3d3d3}\textbf{0.4457} & 0.4672 & A4\\
        		& $f_4(x)$ & 0.2011 & 0.2014 & 0.2291 & 0.1979 & \cellcolor[HTML]{d3d3d3}\textbf{0.1907} & A5\\
        		\bottomrule
            \end{tabular}
        }
	\end{table*}
\end{small}

Considering that IaaS providers may have to deal with several different workloads when provisioning resources to customers with very heterogeneous types of requirements, the average behavior of the compared heuristics could be a relevant information.  This average costs of $f(x)$ are considered to build a ranking of algorithms according to the experiments performed in this work.

To better understand the presented comparison, Table \ref{tab:OFvaluesPerAlgorithm} details $\hat{f_i}(x)$to analyze particular preferences of IaaS providers for the optimization of the considered objectives (e.g. power consumption could be more important in hours where the electricity price is higher).

Based on the information presented in Tables \ref{tab:WSvaluesPerAlgorithm} and \ref{tab:OFvaluesPerAlgorithm}, the Main Findings (MF) of the experimental comparison performed in this part are summarized as follows:

\vspace{1mm}

\textbf{MF1:} \textit{There is no evaluated heuristic that can clearly be considered as the best alternative for all objective functions, considered simultaneously}.

\vspace{1mm}

Additionally, none of the evaluated heuristics performed equally well in all 4 experimental workloads (see Table \ref{tab:WSvaluesPerAlgorithm}). Consequently, an heuristic performing good enough in average when considering different types of workloads could be sufficiently convenient.

A more detailed evaluation could be performed to obtain information for conjuntural preferences of IaaS providers (see Table \ref{tab:OFvaluesPerAlgorithm}, where the best heuristic for each objective function and each workload trace is highlighted in the last column).

\vspace{2mm}

\textbf{MF2:} \textit{FFD heuristic (A4) was the algorithm that outperformed all of the other evaluated heuristics taking into account average results in performed experiments}.

\vspace{1mm}

According to the average performance of each evaluated heuristic (see Table \ref{tab:WSvaluesPerAlgorithm}), the following ranking was built: (1st) FDD (A4), (2nd) BFD (A5), (3th) BF (A2), (4th) FF (A1), and (5th) WF (A3), where BFD follows very close the average performance of FDD (with a difference of 1.75\%).

\vspace{2mm}

\textbf{MF3:} \textit{WF algorithm (A3) is suggested for workloads that can be considered stable (i.e. no workload peaks).}

\vspace{1mm}

For experimental workload trace $W_4$ that considers an Uniform PDF for VM request arrivals, A3 clearly presented the best performance obtaining minimum average results in Table \ref{tab:WSvaluesPerAlgorithm}. It also performed as the best algorithm for 3 out of the 4 objectives, considering only $W_4$ (see Table \ref{tab:OFvaluesPerAlgorithm}).

\vspace{2mm}

\textbf{MF4:} \textit{The WF (A3), BFD (A5), FFD (A4) and BFD (A5) algorithms are recommended for $f_1(x)$, $f_2(x)$, $f_3(x)$ and $f_4(x)$ objective functions respectively, when there is a preferred objective function (lexicographic order).} 

\vspace{1mm}

The BFD algorithm (A5) performed as the best algorithm on average for $f_2(x)$ and $f_4(x)$. For $f_1(x)$ the best alternative seems to be A3 while A4 could be considered as the best for $f_3(x)$ (see Table \ref{tab:OFvaluesPerAlgorithm}).

\vspace{2mm}

\textbf{MF5:} \textit{As expected, the offline MA (A6) outperformed all evaluated online heuristics in all experimental workloads}. 

\vspace{1mm}

An offline MA was compared to the five evaluated heuristics to experimentally demonstrate that online decisions made along a simulation affects the quality of solutions. Clearly, offline algorithms such as the considered MA present a substantial advantage over online heuristics when considering the quality of solutions for the combined objective function $f(x)$ (see Table \ref{tab:WSvaluesPerAlgorithm}). This advantage is presented for the following two main reasons: (1) offline algorithms have a complete knowledge of the future events of a problem instance and (2) only A6 considered migrations of VMs between PMs in the comparison, reconfiguring the placement when convenient.

Having a complete knowledge of future VMs requests is considered unrealistic for IaaS environments \cite{beloglazov2012optimal}. Consequently, online algorithms, such as the heuristics evaluated in this work, are a good alternative for VMP problems in IaaS environments. The evaluated offline MA (A6) improves the quality of solutions with migrations of VMs between PMs, at the expense of costs associated to placement reconfigurations, as seen in Table \ref{tab:memeticMigrationData} where the average VMs migration overhead of each workload is presented as: (1) total number of VM migrations and (2) total RAM memory migrated. 

\vspace{2mm}

\begin{table}[!t]
	\caption{VMs Migration Overhead for MA (A6).}
	\label{tab:memeticMigrationData}
	
	\ra{1.3}
	\centering
	
	\begin{tabular}{@{} l*{2}{r} @{}}
		\toprule
		& \multicolumn{2}{c}{\textbf{Migration}}\\
		\cmidrule(l){2-3}
	    \textbf{Workload} & \textbf{\# of VMs} & \textbf{Memory in [GB]}\\
	    \midrule
		
		Poisson {$\lambda$ = 10} & 1493.7 & 4865.9 \\
		Poisson {$\lambda$ = 50} & 740.0 & 2775.1 \\
		Poisson {$\lambda$ = 70} & 643.6 & 2415.5 \\
		Uniform & 1416.7 & 5329.8 \\
		
		\bottomrule
	\end{tabular}
\end{table}

\begin{table}[ht!]
	\caption{Execution Time per Algorithm in [ms].}
	\label{tab:exec_time}
	
	\ra{1.3}
	\centering
	
	\begin{tabular}{@{} l*{6}{c} @{}}
		\toprule
		& \multicolumn{6}{c}{\textbf{Algorithm}}\\
		\cmidrule{2-7}
		\textbf{Workload} & \textbf{A1} & \textbf{A2} & \textbf{A3} & \textbf{A4} & \textbf{A5} & \textbf{A6} \\
		\midrule
		
		Poisson {$\lambda$ = 10} & 3 & 3 & 3 & 5 & 5 & 55770 \\
		Poisson {$\lambda$ = 50} & 1 & 2 & 2 & 3 & 3 & 31219 \\
		Poisson {$\lambda$ = 70} & 1 & 1 & 1 & 2 & 2 & 23885 \\
		Uniform & 2 & 2 & 3 & 3 & 3 & 44477 \\
		
		\bottomrule
	\end{tabular}
\end{table}

\textbf{MF6:} \textit{As expected, the execution time of all online heuristics were considerably shorter than the offline alternative}.

Considering the low complexity of the evaluated heuristics, these algorithms are able to find promising solutions in short periods of time (a main concern for IaaS providers solving VMP problems) at the expense of not exploring solutions that can potentially result in better quality solutions.

Clearly, any offline alternative that intelligently explore a large space of feasible solutions will result in higher execution times, mainly considering the exponential computational complexity of the VMP problem. (see Table \ref{tab:exec_time})

\chapter{Part I - Conclusions and Future Work}
\label{chap:conclusions_and_future_works_first_part}
This first part of this work presented a formulation to solve a provider-oriented VMP problem optimizing the following objectives: (1) power consumption, (2) economical revenue, (3) quality of service, and (4) resource utilization (\textbf{Objective 1}, published in \cite{lopez2016LANC}). 

An experimental comparison of five different online heuristics against the best performance of an offline MA that had a complete knowledge of future VM requests and VM migration capabilities was performed as part of this work (\textbf{Objective 2}, published in \cite{lopez2016LANC}). A previously studied MA \cite{Ihara2015f} was implemented taking into account that obtaining optimal solutions for large instances of the problem could result impracticable \cite{pires2013multi}. 

Considering the evaluation of the mentioned algorithms solving several experimental workload traces, six main findings (MF1 to MF6) were identified (see Section \ref{sec:heuristicsAnalysis}).

The main advantage of the offline MA (A6) over experimentally evaluated online heuristics (A1 to A5) is a better quality of solutions, as summarized in Table \ref{tab:WSvaluesPerAlgorithm}. 

Unfortunately, offline alternatives such as A6 are not appropriate for dynamic environments of VMP problems for IaaS providers, where future VM requests are unknown and placement should be solved in a very short time.

In order to improve the quality of the solutions obtained by the evaluated heuristics, the VMP problem could be formulated as a two-phase optimization problem. In this context, VMP problems with many objectives for an IaaS could be decomposed into two different sub-problems: (1) incremental VMP (iVMP) and (2) VMP reconfiguration (VMPr) \cite{zheng2015virtual} 
as shown in Figure \ref{fig:two_phase_optimization_scheme}.

A two-phase optimization strategy can be considered, combining both \textit{online} (iVMP) and \textit{offline} (VMPr) algorithms to solve each VMP sub-problem. The iVMP problem is considered for dynamic arriving requests when VMs should be created and removed at runtime. Consequently, this sub-problem should be formulated as an \textit{online} problem and solved in a short period of time, where the studied heuristics could be appropriate. On the contrary, the VMPr problem is considered for improving the quality of solutions obtained by the iVMP, considering placement reconfigurations through migration of VMs. This sub-problem could be formulated \textit{offline}, where alternative solution techniques (e.g. meta-heuristics) could be appropriate.




\vspace*{\fill}
\textbf{
    {\huge
        \begin{center}
        \part{An Experimental Comparison of Scalarization Methods for the Two-Phase Optimization Scheme for the VMP Problem}
        \end{center}
    }
}
\vspace*{\fill}

\lhead{Part II. \emph{An Experimental Comparison of Scalarization Methods for the Two-Phase Optimization Scheme for the VMP Problem}} 

\vspace*{\fill}{
    \begin{center}
        \textcolor{gray}{ \textit{This page is intentionally left blank. }}    
    \end{center}
}
\vspace*{\fill}

\chapter{Part II - Related Work and Motivation}
\label{chap:related_work_and_motivation_second_part}

Currently, IaaS providers manage an enormous amount of resources in order to satisfy clients requests. These resource requests have a highly dynamic behaviour and the IaaS providers try to manage them in such a way that the resources they administrate are used efficiently in order to provide the most economical revenue possible. Resource requests represented as VM placement requests can be grouped into cloud services. It has to be noted that not all of the resources allocated to a specific VM request are used at all times, therefore we have idle allocated resources that can be used elsewhere.

These dynamic resource requests increase the uncertainty of a VMP problem when they are taken into account. In this context, preliminary results help identify that the most relevant dynamic parameters in the VMP literature are \cite{ortigoza2016}: (1) resource capacities of VMs (associated to vertical elasticity) \cite{Li2013a}, (2) number of VMs of a cloud service (associated to horizontal elasticity) \cite{Wang2012} and (3) utilization of resources by VMs (relevant for overbooking) \cite{anand2013virtual}. Considering the mentioned dynamic parameters, environments for IaaS formulations of provider-oriented VMP problems could be classified by one or more of the following classification criteria: (1) service elasticity and (2) overbooking of physical resources \cite{ortigoza2016}




In order to attend the current demand a cloud service could request additional resources to scale up (adding more resources to the existing VMs) or to scale out (adding more VMs to the service), this is characteristic behaviour of a elastic cloud service, therefore IaaS providers should model these requirements accordingly. From an IaaS provider perspective, elastic cloud services should be considered more important than non-elastic ones \cite{jortigoza2015}
. Different IaaS environments could be formulated considering one of the following service elasticity values: no elasticity, horizontal elasticity, vertical elasticity or both horizontal and vertical elasticity \cite{ortigoza2016}.


Additionally, resources of VMs are dynamically used, giving space to re-utilization of idle resources that were already reserved. In this context, IaaS environments identified in \cite{ortigoza2016} may also consider one of the following overbooking values: no overbooking, server resources overbooking, network resources overbooking or both server and network overbooking.

Uncertainty factors like elasticity and overbooking influence the formulation of VMP problems and must not be ignored when a realistic formulation is being proposed. Obtaining scenarios that portray said uncertainty factors is an important aspect when comparing solution techniques of VMP problem. Consequently, workload trace generators \cite{jortigoza2015} must take into consideration the different possible elasticity and overbooking scenarios.


This work formulates a VMP problem taking into account the most complex environment identified in \cite{ortigoza2016}, that considers both types of service elasticity and both types of overbooking of physical resources.

\section{Two-phase Optimization Scheme for the VMP Problem}
\label{sec:two-phase-optimization-scheme}

From an IaaS provider perspective, the VMP is mostly formulated as an online problem and must be solved with short time constraints \cite{lopez2015}. A VMP problem formulation is considered to be online when no knowledge of future VMP requests are modelled, so that its solution technique (e.g. heuristics) dynamically makes placement decisions considering only the requests known up until the time the decision is made \cite{borodin2005online}. On the other hand, if the formulation includes a list of all the VMP requests including future requests, in other words, the solution technique has to solve a VMP problem considering a static environment where VM requests are known in its entirety before execution, then the VMP problem formulation is considered to be offline \cite{borodin2005online}.

Considering the on-demand model of cloud computing with dynamic resource provisioning and dynamic workloads of cloud applications \cite{Mell2011}, the resolution of VMP problems should be performed as fast as possible in order to be able to support these dynamic requirements.

Online decisions made along the operation of a cloud computing infrastructure may negatively affect the quality of obtained solutions when compared to offline decisions \cite{lopez2016LANC}. Unfortunately, offline formulations are not appropriate in highly dynamic environments for real-world IaaS providers where cloud service's allocation is requested dynamically.

To improve the solution's quality obtained by online algorithms, the VMP problem could be formulated as a two-phase optimization problem, combining advantages of online and offline formulations \cite{lopez2016LANC}. In this context, VMP problems could be decomposed in two different sub-problems: (i) incremental VMP (iVMP) and (ii) VMP reconfiguration (VMPr).

The iVMP sub-problem is considered to allocate dynamic arriving requests, where VMs could be created, modified and removed at runtime. Existing heuristics could be reasonably appropriate \cite{lopez2016LANC} for this work.

The VMPr sub-problem is considered to improve the quality of solutions obtained in the iVMP phase, reconfiguring the placement through VM migration, where alternative solution techniques could result more suitable (e.g. meta-heuristics) \cite{lopez2016LANC}.

The iVMP phase is considered at each time instant, while the VMPr phase is triggered according to the considered VMPr triggering method (e.g. periodically). Once the VMPr is triggered, the placement of VMs is recalculated and the iVMP proceeds normally.

Both the iVMP and the VMPr phases must deal with a problem that may affect gravely its performance, this problem entails possible solutions selection from a set of feasible solutions. In Pareto based algorithms, the Pareto set approximation can include a large number of non-dominated solutions; therefore, in a dynamic environment, selecting one of the non-dominated solutions can be considered as a new difficulty for the VMP.

The first part of this work considers a strategy to deal with large numbers of non-dominated solutions, by normalizing their values to finally scalarize them into a single comparable value. Effectively converting the multi-objective optimization problem into a multi-objective as mono-objective optimization problem based on \cite{Ihara2015f}. Thus, reducing the complexity of the solution selection algorithm applied in said part.

There are other methods to scalarize OFs that can be considered as alternatives to the Weighted Sum method, used in the first part.

This work proposes the utilization of three scalarization methods as part of a experimental evaluation to determine which one is the most suitable method to be applied as part of the solution comparison process of the iVMP and VMPr phases in order to improve the performance of the two-phase optimization scheme.


\section{Workload Trace Generation}
\label{sec:workload-trace-generation}
Considering the highly dynamic customers's requests of IaaS resources, the study of VMP problem solution's approaches must take into consideration several aspects of the environment in order to provide close enough approximations to the real behaviour of VMP requests. These aspects, such as elastic services, overbooking of resources and the combination of the previous two \cite{jortigoza2015}, have to be modeled as workload traces to be the input of the possible solutions of VMP problems in order test their capabilities.

Given the dynamic aspect of the VMP problem, workload traces must be modelled as input of online problems, taking into consideration dynamic resources provisioning. Manual generation of workload traces that mimic the real-environment's rapid changes is a not an easy task and a way to systematically generate cloud workload traces can decrease the time to set up VMP problem experiments and help improve the comparison of solution techniques that may use a common input generated with controlled parameters.

In this part of the work, a cloud workload trace generator is developed to provide parametric dynamic input to the solution approaches of the VMP problem that are studied.

\section{Motivation}
\label{sec:motivation_second_part}
Due to the randomness of customer requests, VMP problems should be formulated under uncertainty. This work presents a scenario-based uncertainty approach for modeling uncertain parameters, considering the developed workload trace generator to create input data and considering a two-phase optimization scheme for VMP problems in the proposed complex IaaS environment.

The second part of this work summarizes an experimental evaluation under uncertainty of the three scalarization methods developed as part of the solution comparison process of the iVMP phase and the VMPr phase using generated dynamic cloud workload traces in order to improve the average quality of placement solutions, to answer the following question: \textit{Which scalarization methods are most suitable for solving the two-phase optimization scheme for VMP problems in federated-clouds considering many objectives, elastic and overbooked services ?}

The following chapters summarize the complex IaaS environment for VMP problems considered in this second part, as well as formal definitions of both iVMP and VMPr problems.

\chapter{Two-Phase Optimization Scheme for VMP problem formulation}
\label{chap:problem_formulation_second_part}

The proposed formulation of the VMP problem models a complex IaaS environment, composed by available PMs and VMs requested at each discrete time $t$, considering the following input data:

\begin{itemize}
    \item a set of $n$ available PMs and their specifications (Matrix \ref{matrix:H});
    
    \vspace{-2mm}
    
    \item a set of $m(t)$ VMs requested, at each discrete time $t$, and their specifications (Matrix \ref{matrix:V_t});
    
    \vspace{-2mm}
    
    \item information about the utilization of resources of each active VM at each discrete time $t$ (Matrix \ref{matrix:U_t});
    
    \vspace{-2mm}
    
    \item the current placement at each discrete time $t$ (i.e. x(t)) (Matrix \ref{matrix:P_t}).
\end{itemize}

The proposed iVMP and VMPr sub-problems consider different sub-sets of the above mentioned input data, as later presented in Sections \ref{sec:ivmp} and \ref{sec:vmpr}. The set of PMs owned by the IaaS provider is represented as a matrix $H \in \mathbb{R}^{n\times(r+2)}$, as presented in (Matrix \ref{matrix:H}). Each PM $H_i$ is represented by $r$ different physical resources. This work considers $r=3$ physical resources ($Pr_1$-$Pr_3$): CPU [ECU], RAM [GB] and network capacity [Mbps]. The maximum power consumption [W] is also considered. It is important to mention that the proposed notation is general enough to include more characteristics associated to physical resources such as Graphical Processing Unit (GPU), storage capacity or number of cores, to cite a few. 

Finally, considering that an IaaS provider could own more than one cloud datacenter, PMs notation also includes a datacenter identifier $c_i$, i.e.

\vspace{-2mm}
\begin{equation}
\label{matrix:H}
\hspace{0.3cm}
  H=
  \left[
  \begin{array}{ccccc}
		Pr_{1,1} & \dots & Pr_{k,1} & pmax_1 & c_1\\
		\dots  & \dots  & \dots  & \dots & \dots  \\
		Pr_{1,n} & \dots & Pr_{k,n} & pmax_n & c_n\\
  \end{array}
  \right]
\end{equation}

\vspace{-2mm}
\noindent \textit{where:}

\vspace{-2mm}
\begin{tabbing}
\hspace{2.5cm}\=\kill 
    $i$:        \>PM identifier ($1 \leq i \leq n$);\\
    $Pr_{k,i}$  \>Physical resource $k$ on $H_i$ ($1 \leq k \leq r$); \\
    $pmax_i$: 	\>Maximum power consumption of $H_i$ in [W];\\
	$c_{i}$:    \>Datacenter identifier of $H_{i}$, where $1 \leq c_i \leq c_{max}$;\\
    $n$:      	\>Total number of PMs.
\end{tabbing}

Clearly, the set of PMs $H$ could be modeled as a function of time $t$, considering PM crashes \cite{sv2015continuous}, maintenance or even deployment of new hardware. The mentioned modeling approach for PMs is out of the scope of this work and its particular considerations are left as future work (see Chapter \ref{chap:conclusions_and_future_works_second_part}). 

In the complex environment considered in this work, an IaaS provider dynamically receives requests (i.e. a set of inter-related VMs) at each discrete time $t$. A cloud service $S_b$ is composed by a set of VMs which may be placed in different datacenters according to a customer preference or requirement.

The set of VMs requested by customers at each discrete time $t$ is represented as a matrix $V(t) \in \mathbb{R}^{m(t)\times(r+2)}$, as presented in (Matrix \ref{matrix:V_t}). In this work, each VM $V_{j}$ requires $r=3$ (according to PM $H$'s physical resources) different virtual resources ($Vr_1$-$Vr_3$): CPU [ECU], RAM [GB] and network capacity [Mbps]. Additionally, a cloud service identifier $b_j$ is considered, as well as an economical revenue $R_j$ [USD] associated to each VM $V_{j}$, which is the economical cost that a customer has to pay to the IaaS provider for hosting the VM $V_{j}$.

As mentioned before, the proposed notation could represent any other set of resources. The requested VMs try to lease virtual resources for an unknown period of discrete time.

\vspace{-2mm}
\begin{equation}
\label{matrix:V_t}
\setlength{\arraycolsep}{1pt}
\hspace{-0.5cm}
 	V(t)= \left[
		\begin{array}{ccccc}
			Vr_{1,1}(t) & \dots & Vr_{k,1}(t) & b_1 & R_1(t) \\
			\dots & \dots & \dots & \dots & \dots \\
      Vr_{{1,m(t)}}(t) & \dots & Vr_{{k,m(t)}}(t) & b_m(t) & R_{m(t)}(t)
    \end{array}
	\right]
\end{equation}

\vspace{-2mm}
\noindent \textit{where:}

\vspace{-2mm}
\begin{tabbing}
\hspace{2.5cm}\=\kill
    $j$:            \>VM identifier ($1 \leq j \leq m(t)$);\\
	$Vr_{k,j}(t)$:  \>Virtual resource $k$ on $V_j$, where $1 \leq k \leq r$;\\
	$b_{j}$:        \>Service identifier of $V_{j}$;	\\
	$R_j(t)$:       \>Economical revenue for allocating $V_j$ in [USD]; \\
	$m(t)$:         \>Number of VMs at each discrete time $t$, where $1 \leq m(t) \leq m_{max}$; \\
	$m_{max}$:      \>Maximum number of VMs; \\
\end{tabbing}

Once a VM $V_{j}$ is powered-off by a customer, its virtual resources are released, so the IaaS provider can reuse them. For simplicity purposes, in what follows the index $j$ is not reused.

In order to model a dynamic VMP environment taking into account both vertical and horizontal elasticity of cloud services \cite{jortigoza2015}, the set of requested VMs $V(t)$ may include the following types of requests at each time $t$: 

\begin{itemize}
	\item \textbf{cloud service creation:} where a new cloud service $S_b$, composed by one or more VMs $V_j$, is created. Consequently, the number of VMs at each discrete time $t$ (i.e. $m(t)$) is a function of time;
	
	\vspace{-2mm}
	\item \textbf{scale-up / scale-down of VMs resources:} where one or more VMs $V_j$ of a cloud service $S_b$ increases (scale-up) or decreases (scale-down) the amount of allocated virtual resources considering the current demand (vertical elasticity). In order to model these considerations, virtual resource capacities of a VM $V_j$ (i.e. $Vr_{1,j}(t)$-$Vr_{3,j}(t)$) are time functions, as well as the associated economical revenue ($R_j(t)$);
	
	\vspace{-2mm}
	\item \textbf{cloud service scale-out / scale-in:} where a cloud service $S_b$ increases (scale-out) or decreases (scale-in) the number of associated VMs according to current demand (horizontal elasticity). Consequently, the number of VMs $V_{j}$ in a cloud service $S_{b}$ at each discrete time $t$ is a function of time;
	
	\vspace{-2mm}
	\item \textbf{cloud service destruction:} where virtual resources of cloud services $S_b$, composed by one or more VMs $V_j$, are released.
\end{itemize}

In most situations, virtual resources requested by cloud services are dynamically used, giving space to re-utilization of idle resources that were already reserved. Information on the utilization of virtual resources at each discrete time $t$ is required in order to model a dynamic VMP environment where IaaS providers consider overbooking of, both server and networking, physical resources.

Resource utilization of a VM $V_{j}$ at each discrete time $t$ is represented as a matrix $U(t) \in \mathbb{R}^{m(t) \times r}$:

\vspace{-2mm}
\begin{equation}
    \label{matrix:U_t}
 	U(t)= \left[
		\begin{array}{ccc}
			Ur_{1,1}(t) & \dots & Ur_{k,1}(t) \\
			\dots & \dots & \dots \\
      Ur_{{1,m(t)}}(t) & \dots & Ur_{{k,m(t)}}(t)
    \end{array}
	\right]
\end{equation}

\vspace{-2mm}
\noindent \textit{where:}

\vspace{-2mm}
\begin{tabbing}
\hspace{2.5cm}\=\kill 
    $Ur_{k,j}(t)$   \>is the utilization ratio of $Vr_{k}(t)$ in $V_j$ at each discrete time $t$.
\end{tabbing}

The current placement of VMs into PMs $(x(t))$ considers VMs requested in the previous discrete time $t-1$. Consequently, the dimension of $x(t)$ is based on the number of VMs $m(t-1)$. Formally, the placement at each discrete time $t$ is represented as a matrix $x(t) \in \{0,1\}^{m(t-1) \times n}$, as defined in (Matrix \ref{matrix:P_t}):

\vspace{-2mm}
\begin{equation} \label{matrix:P_t}
	x(t)= \left[
	\begin{array}{cccc}
		x_{1,1}(t) & x_{2,1}(t) & \dots & x_{n,1}(t) \\
		\dots & \dots & \dots & \dots \\
		x_{1,m(t-1)}(t) & x_{2,m(t-1)}(t) & \dots & x_{n,m(t-1)}(t)
	\end{array}
\right]
\end{equation}

\vspace{-2mm}
\noindent \textit{where:}

\vspace{-2mm}
\begin{tabbing}
\hspace{2.7cm}\=\kill
    $x_{j,i}(t) \in \{0,1\}$:   \>indicates if $V_j$ is allocated $(x_{j,i}(t) = 1)$ or not $(x_{j,i}(t) = 0)$ for execution\\
                                \> in a PM $H_i$ at a discrete time $t$ (i.e., $x_{j,i}(t) : V_j \rightarrow H_i)$.
\end{tabbing}

\section{Incremental VMP (iVMP)}
\label{sec:ivmp}

In online algorithms for solving the proposed VMP problem placement decisions are performed at each discrete time $t$ without knowledge of upcoming VM requests. The formulation of the proposed iVMP (online) problem is based on \cite{lopez2016LANC} and could be formally enunciated as:

\noindent \textit{Given an advanced dynamic IaaS environment composed by a set of PMs $(H)$ and a set of VMs requested at each discrete time $t$ $(V(t))$, as well as the current placement of VMs into PMs (P(t)), it is sought an incremental placement of $V(t)$ into $H$ for the discrete time $t+1$ $(P(t+1))$ without migrations, satisfying the problem constraints and optimizing the considered objective functions.}


\subsection{Input Data for iVMP}
\label{subsec:input_data_ivmp}

The proposed formulation of the iVMP problem receives the following information as input data:

\begin{itemize}
	
	\item a set of $n$ available PMs and their specifications (Matrix \ref{matrix:H});
	
	\vspace{-2mm}
	\item a set of $m(t)$ VMs requested at each discrete time $t$ and their specifications (Matrix \ref{matrix:V_t});
	
	\vspace{-2mm}
	\item information about the utilization of resources of each active VM at each discrete time $t$ (Matrix \ref{matrix:U_t});
	
	\vspace{-2mm}
	\item the current placement at each discrete time $t$ (i.e. $x(t)$, Matrix \eqref{matrix:P_t}).
\end{itemize}

\subsection{Output Data for iVMP}
\label{subsec:output_data_ivmp}

The result of the iVMP phase at each discrete time $t$ is an incremental placement $\Delta x$ for the next time instant in such a way that $x(t+1) = x(t) + \Delta x$. The placement at $t+1$ is represented as the matrix $x(t+1) \in \{0,1\}^{n \times m(t)}$, as defined in (Matrix \ref{matrix:P_t1}):

\vspace{-2mm}
\begin{equation}
    \label{matrix:P_t1}
    x(t+1)= \left[ 
        \begin{array}{cccc}
		    x_{1,1}(t+1) & x_{2,1}(t+1) & \dots & x_{n,1}(t+1) \\
		    \dots & \dots & \dots & \dots \\
		    x_{1,m(t)}(t+1) & x_{2,m(t)}(t+1) & \dots & x_{n,m(t)}(t+1)
        \end{array}
    \right]
\end{equation}

Formally, the incremental placement for the next time instant $x(t+1)$ is a function of the current placement $x(t)$ and the VMs requested at discrete time $t$ $V(t)$, defined as:

\vspace{-2mm}
\begin{equation}
    \label{eq:x_t1}
    x(t+1) = f\left[x(t),V(t)\right]
\end{equation}

\section{VMP Reconfiguration (VMPr)}
\label{sec:vmpr}

Previous research work by the authors focused on developing VMPr algorithms considering centralized decisions such as the offline MAs presented in \cite{Lopez-Pires2013, Ihara2015f, lopez2015b}. An offline algorithm has knowledge of the complete set of VM requests in order to decide the placement of these VMs into available PMs. The formulation of the proposed VMPr (offline) problem is based on \cite{lopez2015b,Ihara2015f} and could be enunciated as:

An offline algorithm solves a VMP problem by migrating VMs between PMs, considering a static environment where VM requests do not change over time. The formulation of the proposed VMPr (offline) problem is based on \cite{Ihara2015f,lopez2015b} and could be enunciated as:

\noindent \textit{Given a current placement of VMs into PMs $(x(t))$, it is sought a placement reconfiguration through migration of VMs between PMs for the discrete time $t$ (i.e. $x'(t)$), satisfying existing constraints and optimizing the considered objective functions.}

\subsection{Input Data for VMPr}
\label{subsec:input_data_vmpr}

The proposed formulation of the VMPr problem receives the following information as input data:

\begin{itemize}
	\item a set of $n$ available PMs and their specifications (Matrix \ref{matrix:H});
	
	\vspace{-2mm}
	\item information about the utilization of resources of each active VM at discrete time $t$ (Matrix \ref{matrix:U_t});
	
	\vspace{-2mm}
	\item the current placement at discrete time $t$ (i.e. $x(t)$, Matrix \ref{matrix:P_t}).
\end{itemize}

\subsection{Output Data for VMPr}
\label{subsec:output_data_vmpr}

\begin{figure}[!ht]
	\centering
	\includegraphics[scale=0.16]{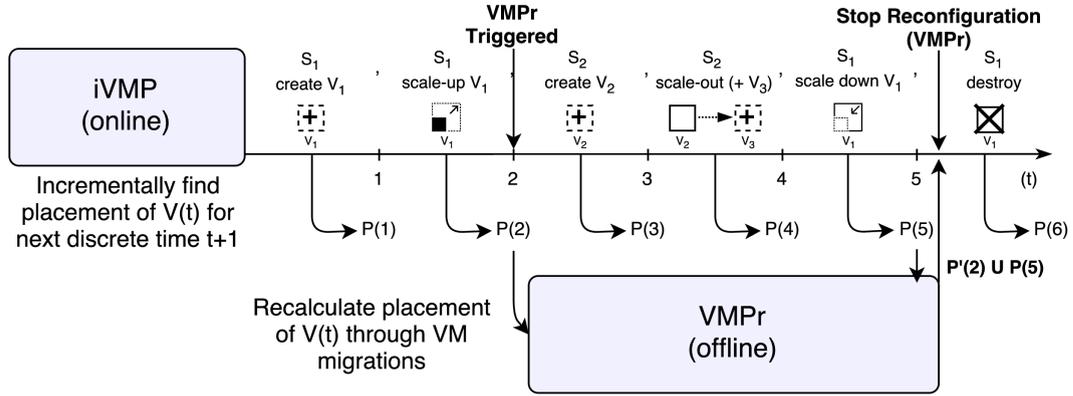}
	\caption{Two-Phase optimization scheme for VMP problems.}
	\label{fig:two_phase_optimization_scheme}
\end{figure}

The result of the VMPr problem is a placement reconfiguration through migration of VMs between PMs for the discrete time $t$, represented by a placement reconfiguration of $x(t)$ (i.e. $x'(t)$, Matrix \ref{matrix:P_t}), which represents the output of the VMPr process. Logically, this new placement $x'(t)$ should be upgraded with requests arrived during VMPr calculation before migrating VMs and stopping reconfiguration, as shown in Figure \ref{fig:two_phase_optimization_scheme}.


\section{The Two-Phase Optimization Process}
\label{sec:two_phase_optimization_process}

Figure \ref{fig:two_phase_optimization_scheme} illustrates the two-phase optimization scheme process for VMP problems. The iVMP phase is considered at each instant $t$ to attend the set of requested VMs, that may include the following types of requests: 
cloud service creation, scale-up/scale-down (vertical-elasticity), cloud service scale-out/scale-in (horizontal-elasticity) and
cloud service destruction (details were described above). The iVMP phase dynamically allocates requested VMs and finds the placement of $V(t)$ for the next discrete time $t+1$, the obtained placement may be not optimal against an offline solution as concluded in Chapter \ref{chap:conclusions_and_future_works_first_part}.

In order to improve the solution's quality, the VMPr phase is triggered according to the considered VMPr triggering method (VMPr triggered at $t=2$). It takes into account the current placement of the datacenter (in this case $P(2)$) and recalculates the placement of $V(t)$ through VMs migration. While the VMPr is running, the iVMP keeps attending the set of requested VMs at each instant $t$.

Once the VMPr is done, the obtained placement by the VMPr ($P'(2)$) doesn't have all the requests (requests kept arriving during execution), 
in view of that, the VMPr joins the $P'(2)$ and the current placement $P(5)$. Finally if the resulting placement is considered to be better than the current placement, it replaces the current placement of the two-phase optimization scheme, as shown in Figure \ref{fig:two_phase_optimization_scheme} at $t=6$; otherwise, the resulting placement is discarded.

\section{Constraints}
\label{sec:constraints_part_2}

Several constraints should be considered when solving a VMP. For example, a VM $V_{j}$ should be allocated at a provider's datacenter or alternatively be derived to a federated IaaS partner, the VM $V_{j}$ with the highest level of SLA or economical revenue $R$ must be allocated to run on a PM $H_i$, to cite a few constraints.

\subsection{Constraint 1: Unique Placement of VMs}
\label{subsec:constraint_1_second_part}
A VM $V_{j}$ should be allocated to run on a single PM $H_i$ or alternatively derived in a federated IaaS partner. From an IaaS provider's perspective, elastic cloud services are usually considered more important than non-elastic ones \cite{jortigoza2015}. Consequently, resources of elastic cloud services are most of the time assigned a higher priority than non-elastic ones, usually reflected in legal contracts. This placement constraint is expressed as:

\vspace{-2mm}
\begin{equation}\label{eq:unique_placement_second_part}
	\begin{aligned}
		&\sum_{i=1}^{n} x_{j,i}(t) \leq 1
	\end{aligned}
\end{equation}

\vspace{-2mm}
\begin{center}
	$\forall j \in \{1,\dots,m(t)\}, \text{i.e. for all VM $V_j$}$.
\end{center}

\vspace{-2mm}
\noindent \textit{where:}
\vspace{-2mm}
\begin{tabbing}
\hspace*{3cm}\=\kill
        $x_{j,i}(t) \in \{0,1\}$:   \>Indicates if $V_j$ is allocated $(x_{j,i}(t) = 1)$ or not $(x_{j,i}(t) = 0)$\\
                                    \> for execution in a PM $H_i$ (i.e., $x_{j,i}(t) : V_j \rightarrow H_i)$ at a discrete time $t$; \\
        $n$:                        \>Total number of PMs; \\
        $m(t)$:                     \>Number of VMs at each discrete time $t$, where $1 \leq m(t) \leq m_{max}$.\\
\end{tabbing}

\subsection{Constraints 2-4: Overbooked Resources of PMs}
\label{subsec:constraint2-4_second_part}

At the same time, a PM $H_i$ must have sufficient available resources to meet the dynamic requirements of all VMs $V_{j}$ that are allocated to run on $H_i$. It is important to remember that resources of VMs are dynamically used, giving space to re-utilization of idle resources that were already reserved. Re-utilization of idle resources could represent a higher risk of unsatisfied demand in case of peak resources' utilization over a short period of time. Therefore, providers need to reserve a percentage of idle resources as a protection, defined by a overbooking protection factor ($\lambda_k$). This protection factor sets the amount of resources to reserve in order to mitigate SLA violations and can take values between 0, meaning overbooking of all the available resources, and 1, no overbooking of resources at all. These constraints can be formulated as:

\vspace{-2mm}
\begin{equation}\label{eq:phys_cpu_second_part}
	\begin{aligned}
		\sum_{j=1}^{m(t)} \{Vr_{k,j} \times Ur_{k,j} + [Vr_{k,j} \times (1 - Ur_{k,j})] \times \lambda_k\} \times x_{j,i}(t) \le Pr_{k,i}
	\end{aligned}
\end{equation}

\begin{center}
$\forall i \in \{1,\dots,n\}$ and $\forall k \in \{1,\dots,r\}$, \\
i.e. for each PM $H_i$ and for each considered resource $r$.
\end{center}

\vspace{-2mm}
\noindent \textit{where:} 
\vspace{-2mm}
\begin{tabbing}
\hspace*{2.5cm}\=\kill
    $\lambda_k$:        \>is the protection factor for $Vr_{k,j}$ $\in$ [0,1].
\end{tabbing}

Physical resources are considered as resources entirely available to VMs, without considering resources for the PM's hypervisor.

\section{Objective Functions}
\label{sec:objective_functions_second_part}

Each of the considered objective functions must be formulated in a single optimization context (i.e. minimization or maximization) and each objective function cost must be normalized to be comparable and combinable as a single objective. This work normalizes each objective function cost by calculating $\hat{f_i}(x) \in \mathbb{R}$, where $0 \leq \hat{f_i}(x) \leq 1$.

\subsection{Power Consumption Minimization}
\label{subsec:powerConsumption_second_part}
Based on Beloglazov et al. \cite{beloglazov2012energy}, this work models the power consumption of PMs considering a linear relationship with the CPU utilization of PMs, without taking into account PMs at alternative datacenters of the cloud federation. The power consumption minimization can be represented by the sum of the power consumption of each PM $H_i$ that composes the advanced IaaS environment (see Chapter \ref{chap:problem_formulation_second_part}), as defined in Equation (\ref{eq:min_power_consumption_second_part}).

\vspace{-2mm}
\begin{equation}  \label{eq:min_power_consumption_second_part}
	f_1(t) = \sum_{i=1}^{n} ((pmax_i - pmin_i) \times Ur_{1,i}(t) + pmin_i) \times Y_i(t)
\end{equation}

\vspace{-2mm}
\noindent \textit{where:}

\vspace{-2mm}
\begin{tabbing}
\hspace*{3cm}\=\kill
	$f_1(t)$:               \>Total power consumption of PMs; \\
	$n$:                    \>The total number of PMs; \\
	$pmax_i$:               \>Maximum power consumption of a PM $H_i$; \\
	$pmin_i$:               \>Minimum power consumption of a PM $H_i$. As suggested in \cite{beloglazov2012energy}, \\
                            \> $pmin_i \approx pmax_i * 0.6$; \\
    $Ur_{1,i}(t)$:          \>Utilization ratio of resource 1 (in this case CPU) by $H_i$ at instant $t$;\\
	$Y_i(t) \in \{0,1\}$:   \>Indicates if $H_i$ is turned on $(Y_i(t) = 1)$ or not $(Y_i(t) = 0)$ at instant $t$.
\end{tabbing}

\subsection{Economical Revenue Maximization}
\label{subsec:economicalRevenue_part_2}

For IaaS customers, cloud computing resources often appear to be unlimited and can be provisioned in any quantity at any required time $t$ \cite{Mell2011}. Consequently, this work considered a basic federated-cloud deployment architecture, where a main provider may attend requested resources that are not able to be provided (e.g. a workload peak) by transparently leasing low-price resources from alternative datacenters owned by federated providers \cite{6779415}. This leasing costs should be minimized in order to maximize economical revenue objective function.

Equation (\ref{eq:leasing_cost}) represents the mentioned leasing costs, defined as the sum of the total costs of leasing each VM $V_j$ that is effectively allocated for execution on any PM of an alternative datacenter of the cloud federation. A provider must offer its idle resources to the cloud federation at lower prices than offered to customers in the actual cloud market. The pricing scheme may depend on the particular agreement between providers of the cloud federation \cite{6779415}. For simplicity, this work considers that the main provider may lease requested resources (that are not able to provide) from the cloud federation at 70\% $(\hat{X_j} = 0.7)$ of its price in markets ($R_j(t)$). These costs may be formulated as:

\vspace{-2mm}
\begin{equation} \label{eq:leasing_cost}
	\text{Leasing Costs } = \sum_{j=1}^{m(t)} (R_j(t) \times X_j(t) \times \hat{X_j})
\end{equation}

\vspace{-2mm}
\noindent	\textit{where:}

\vspace{-2mm}
\begin{tabbing}
\hspace*{3cm}\=\kill
	$R_j(t)$:               \>Economical revenue for attending $V_j$ in [USD]; \\
	$X_j(t)\in \{0,1\}$:    \>Indicates if $V_j$ is allocated for execution on a PM $(X_j(t) = 1)$ or\\
                            \> not $(X_j(t) = 0)$ at instant $t$; \\
	$\hat{X_j}$:            \>Indicates if $V_j$ is allocated on the main provider $(\hat{X_j} = 0)$ or\\
                            \> on an alternative datacenter of the cloud federation $(\hat{X_j} = 0.7)$;\\
	$m(t)$:                 \>Number of VMs at each discrete time $t$, where $1 \leq m(t) \leq m_{max}$.					
\end{tabbing}

It is important to note that $\hat{X_j}$ is not a function of time. The decision of locating a VM $V_j$ on a federated provider is considered only in the placement process, with no possible migrations between providers. The value of $\hat{X_j}$ depends on the agreement celebrated between federated providers.

Additionally, overbooked resources may inccur in unsatisfied demand of resources at some periods of time, causing QoS degradation, and consequently SLA violations with economical penalties. This economical penalties should be minimized for an economical revenue maximization.

Based on the workload independent QoS metric presented in \cite{beloglazov2012energy}, formalized in SLAs, this work presents Equation (\ref{eq:ud}) to represent total economical penalties for SLA violations, defined as the sum of the total proportional penalties costs for unsatisfied demand of resources.

\vspace{-2mm}
\begin{equation} \label{eq:ud}
	\text{Economical Penalties } = \sum_{j=1}^{m(t)} \bigg[\sum_{q=1}^{r} (Rr_{q,j}(t) \times \Delta r_{q,j}(t) \times X_j(t)\bigg]
\end{equation}

\vspace{2cm}
\noindent	\textit{where:}

\vspace{-2mm}
\begin{tabbing}
\hspace*{3cm}\=\kill
	$r$:                    \>Number of considered resources. In this work 3 (CPU, RAM memory\\
                            \>  and network capacity);\\
	$Rr_{q,j}(t)$:          \>Economical revenue for attending $Vr_{q,j}$; \\
	$\Delta r_{q,j}(t)$:    \>Ratio of unsatisfied resources of the resource $q$ at instant $t$;\\
	$X_j(t)\in \{0,1\}$:    \>Indicates if $V_j$ is allocated for execution on a PM $(X_j(t) = 1)$ or\\
                            \> not $(X_j(t) = 0)$ at instant $t$; \\
	$m(t)$:                 \>Number of VMs at each discrete time $t$, where $1 \leq m(t) \leq m_{max}$.
\end{tabbing}

In this work, the maximization of the total economical revenue that a provider receives for attending the requirements of its customers is achieved by minimizing the total costs of leasing resources from alternative datacenters of the cloud federation as well as minimizing the total economical penalties from SLA violations. 

\vspace{-2mm}
\begin{equation} \label{eq:economical_revenue_second_part}
	f_2(x,t)  = \text{Leasing Costs} + \text{Economical Penalties}
\end{equation}

\vspace{-2mm}
\noindent	\textit{where:}

\vspace{-2mm}
\begin{tabbing}
\hspace*{3cm}\=\kill
	$f_2(x,t)$:       \>Total economical expenditure of the main IaaS provider at instant $t$.
\end{tabbing}

\subsection{Resources Utilization Maximization}
\label{subsec:res_wasted_second_part}
An efficient utilization of resources is a relevant management challenge to be addressed by IaaS providers. This work proposes the resource utilization maximization by minimizing the average ratio of wasted resources on each PM $H_i$ (i.e. resources that are not allocated to any VM $V_j$). This objective function is formulated in Equation (\ref{eq:res_wasted_min_second_part}).

\vspace{-2mm}
\begin{equation}
\label{eq:res_wasted_min_second_part}
    f_3(x, t) = 
		\frac{\sum_{i=1}^{n}
		\bigg[1-
			\bigg(
				\dfrac{\sum_{k=1}^{r} Ur_{k,i}(t)}{r}
			\bigg)
		\bigg] 
		\times Y_i(t)}
		{\sum_{i=1}^{n} Y_i(t)}
\end{equation}

\vspace{-2mm}
\noindent \textit{where:}

\vspace{-2mm}
\begin{tabbing}
\hspace*{3cm}\=\kill
    $f_3(x,t)$:             \>Average ratio of wasted resources; \\
	$Ur_{k,i}(t)$:          \>Utilization ratio of resource $r$ (any resource) by $H_i$ at instant $t$; \\
	$r$:                    \>Number of considered resources. In this work 3 (CPU, \\
                            \> RAM memory and network capacity);\\
    $Y_i(t) \in \{0,1\}$:   \>Indicates if $H_i$ is turned on $(Y_i(t) = 1)$ or not $(Y_i(t) = 0)$ at instant.
\end{tabbing}

\subsection{Reconfiguration Time Minimization}
Performance degradation may occur when migrating VMs between PMs \cite{beloglazov2012optimal}. Logically, it is desirable to keep the time of placement reconfiguration by migration of VMs to the minimum possible. As explained in \cite{beloglazov2012optimal}, the time that a VM takes to be migrated from one PM to another could be estimated as the ratio between the total amount of RAM memory to be migrated and the capacity of the network channel.

In this work, once a placement reconfiguration is triggered, all VM migrations are performed concurrently through a management network 
deployed exclusively for these type of actions, increasing 10\% of CPU utilization in VMs being migrated. Consequently, the reconfiguration time minimization could be achieved by minimizing 
the maximum amount of memory to be migrated from one PM $H_i$ to another $H_i'$ $(i \neq i')$.

Equation (\ref{eq:min_overhead}) is proposed to minimize the maximum amount of RAM memory that must be migrated between PMs at instant $t$.

\vspace{-2mm}
\begin{equation} \label{eq:min_overhead}
	f_4(x,t) = \max (MT_{i,i'}) \quad \forall i,i' \in \{1,\dots,n\}
\end{equation}

\vspace{-2mm}
\noindent \textit{where:}

\vspace{-2mm}
\begin{tabbing}
\hspace*{3cm}\=\kill
    $f_4(x, t)$:    \>Network traffic overhead for VM migrations at instant $t$; \\
    $MT_{i,i'}$:    \>Total amount of RAM memory to be migrated from PM $H_i$ to $H_i'$. 
\end{tabbing}

It should be noted that there are several possible approaches to estimate the migration overhead, as presented in \cite{svard2015principles}.

The following section summarizes the main considerations taken into account to combine the four presented objective functions into a single objective.

\section{Normalization and Scalarization Methods}
\label{sec:norm_and_scalar_part_two}

Given that many objective functions lead to a considerable number of non-dominated solutions and analyzing all of them may consume a large amount of time, a single value that represents a possible solution may greatly decrease the comparison time between solutions, thus increasing the overall solution technique's performance.

Considering the experimental results presented in \cite{Ihara2015f} which supports the resolution of Multi-objective VMP considering a Mono-Objective approach using a scalarization method, each objective function considered as a minimization problem must be normalized to be comparable and combinable into a single objective.

This work normalizes each objective function cost by calculating $\hat{f_i}(x,t) \in \mathbb{R}$, where $0 \leq \hat{f_i}(x,t) \leq 1$, using Equation \eqref{eq:norm_FO_second_part}.



\vspace{-2mm}
\begin{equation}  \label{eq:norm_FO_second_part}
	\hat{f_i}(x, t) = \frac{f_i(x, t) - f_i(x, t)_{min}}{f_i(x, t)_{max} - f_i(x, t)_{min}}
\end{equation}

\vspace{-2mm}
\noindent where:

\vspace{-2mm}
\begin{tabbing}
\hspace*{3cm}\=\kill
	$\hat{f_i}(x, t)$:  \>Normalized cost of objective function $f_i(x, t)$ at instant $t$;\\
	${f_i}(x, t)$:      \>Cost of objective function $f_i(x, t)$;\\
	$f_i(x, t)_{min}$:  \>Minimum possible cost for $f_i(x, t)$;\\
	$f_i(x, t)_{max}$:  \>Maximum possible cost for $f_i(x, t)$.
\end{tabbing}

This work explores different methods that could be applied to a multi-objective optimization to consolidate its normalized objective function into a single comparable value which experimentally yields better results than other alternatives \cite{Ihara2015f}. This work refers to consolidation methods as scalarization methods (SM).

The utilization of the SMs to compare solutions is explained in Sub-Section \ref{subsec:uncertainty}. The three scalarization methods ($F(x,t)$) considered in this work are: (i) Weighted Sum (WS), (ii) Euclidean distance (ED) and (iii) Chebyshev distance (CD). Detailed SM information is presented in Chapter \ref{chap:evaluated_algorithm_second_part}.

\subsection{Scenario-based Uncertainty Modeling}
\label{subsec:uncertainty}

Since this work has required several uncertainty factors to be present in the workload traces. In the second part of this work, uncertainty is modeled through a finite set of well-defined scenarios $S$ \cite{aloulou2008complexity}, where the following uncertain parameters are considered: 

\vspace{-2mm}
\begin{itemize}
    \item virtual resources capacities (vertical elasticity);
    
    \vspace{-2mm}
    \item number of VMs that compose cloud services (horizontal elasticity);
    
    \vspace{-2mm}
    \item utilization of CPU and RAM virtual resources (relevant for overbooking);
    
    \vspace{-2mm}
    \item utilization of networking virtual resources (relevant for overbooking).
\end{itemize}

In order to compare solutions, a representative value per solution must be found taking into consideration the performance of said solution in all of the proposed scenarios.

For each scenario $s \in S$, an average value of the OF's scalarized values $F(x,t)$ is calculated as:

\vspace{-2mm}
\begin{equation}\label{eq:f_s}
	\begin{aligned}
		\overline{f_s(x,t)} = \frac{\sum_{t=1}^{t_{max}} F(x,t)}{t_{max}}
	\end{aligned}
\end{equation}

\vspace{-2mm}
\noindent \textit{where:}

\vspace{-2mm}
\begin{tabbing}
\hspace{3cm}\=\kill
	$\overline{f_s(x,t)}$:  \>The average objective function cost for all discrete time instants $t$\\
	                        \> in scenario $s \in S$;\\
    $F(x,t)$:               \>The scalarized value of the Objective Function values at instant $t$\\
	$t_{max}$:              \>The duration of a scenario in discrete time instants.
\end{tabbing}

As previously described, when parameters are uncertain, it is important to find solutions which considers every scenario $s \in S$. This is achieved by finding the average value of the performance of a solution in every scenario, as defined next in \eqref{eq:avg}.

\vspace{-2mm}
\begin{equation}
\label{eq:avg}
	\begin{aligned}
		\overline{F(x,t)} = \frac{\sum_{s=1}^{|S|} \overline{f_s(x,t)}}{|S|}
	\end{aligned}
\end{equation}

\vspace{-3mm}
\noindent \textit{where:} 

\vspace{-2mm}
\begin{tabbing}
\hspace*{3cm}\=\kill
	$\overline{F(x,t)}$:    \>Average $\overline{f_s(x,t)}$ of a solution for all scenarios $s \in S$; \\
	$|S|$:                  \>Indicates the cardinality of $S$.
\end{tabbing}

\chapter{Cloud Workload Trace Generator}
\label{chap:cwtg}

In order to provide controlled environments for the proposed experiments, a workload trace that simulates the behaviour of real dynamic resource requests is essential. Considering there is no widely known \textit{benchmark} for the VMP problem, a Cloud Workload Trace Generator (CWTG) is developed based on the work presented by Ortigoza et al. in \cite{ortigoza2016CLEI}.

The CWTG considers the resources modelled in the formulation of the problem (Chapter \ref{chap:problem_formulation_second_part}), the uncertainty factors (Sub-Section \ref{subsec:uncertainty}), the different patterns that resource requests can follow in real implementations of datacenters and real configuration of PMs used by IaaS providers (i.e. AWS Instances).

This work proposes to represent a workload trace as a list of VM snapshots. This snapshot list contains a representation of a VM that belongs to a service, this representation contains the detailed list of resources requested by the VM as well as the utilization rates of said resources that change over time.

The VM snapshot list simulates the request types like cloud service creation, scaling up or down of VM resources, cloud service scaling in or out and cloud service destruction. Therefore, the dynamic usage and the uncertainty factors like elasticity and overbooking are well represented in the resulting workload trace.

The CWTG is available online for research purposes \footnote{\small \url{http://github.com/DynamicVMP/workload-trace-generator}}.

\section{Input Data}
\label{sec:cwtg_input}
The generator takes three sets of parameters as input, the general simulation parameters, the elasticity parameters and the overbooking parameters.

The general simulation parameters establish the runtime of the simulation, the number of datacenters, the number of services and VMs per service, it also indicates the PM instance types to be used in the simulation. The elasticity parameters are used to simulate the amount of VMs requested by a service at each time. The overbooking parameters are used to simulate the amount of resources requested by a VM and the utilization of said resources.

The generator implements two PDFs to simulate uncertainty, the Uniform and the Poisson functions. It is important to mention that the generator can be extended in order to support more distribution functions.

\begin{table}[!hb]
	\caption{CTWG Input example}
	\label{tab:input_example_cwtg}
	
	\ra{1.3}
	\centering
	
	\begin{tabular}{@{} ll @{}}
		\toprule
		\textbf{Parameter Description} & \textbf{Input Data} \\
		\midrule
		\multirow{2}{*}{1. Environment} & No Elasticity \\ 
							 			& No Overbooking \\
		\\
		2. Workload Trace Duration [$t$] & 10 \\ 
		3. Number of Cloud Datacenters & 1 \\
		4. Number of Cloud Services & 2 \\
		5. Number of VMs per Service & 5 \\
		7. Instance Types & List of a single instance type\\
		6. Horizontal Elasticity & Uniform(5, 5) \\
  		7. Vertical Elasticity & Uniform(1, 1)\\
		8. Server Resources & Uniform(100, 100)\\
		9. Network Resources & Uniform(100, 100)\\
		\bottomrule
	\end{tabular}
\end{table}

The parameters in Table \ref{tab:input_example_cwtg} will produce a workload trace that lasts 10 units of time, simulating a single datacenter with at most 2 services running simultaneously with no more than 3 VMs per Service. The environment selected for the trace indicates that no elasticity nor overbooking must be considered as an uncertainty factors.

The number of VMs per service follow a Uniform distribution function with the same minimum and maximum values, this way a constant number is ensured to be chosen. The other uncertainty factors also use Uniform distribution functions that will generate a VMs with a static amount of resource requested and a constant utilization of all of it resources.

\section{Workload Trace Generation Process}
\label{sec:cwtg_generation_process}

Basically, the CWTG simulation process follows the pseudo-code shown in Algorithm \ref{alg_main_cwtg}, it initializes all the services setting the starting and an ending time for each. For each unit of time, it retrieves the list of services that are alive at the specific time unit, as visible in line $3$. It iterates over the list of living services from line $4$ to $5$, generating a number of VMs that the service will require at the specific time considering the horizontal elasticity function as seen in line $5$. It generates the VM requests for the each living service in line $6$, this is explained in more detail in Algorithm \ref{alg_vm_generation}. Finally, it prints the snapshots of the each VM request according off all the services ordered by the unit time in which they were on.

\begin{algorithm}[!ht]
\caption{CTWG Algorithm}
\label{alg_main_cwtg}
	\SetKwFunction{Union}{Union}
	\SetKwFunction{UpdatePlacementMatrix}{UpdatePlacementMatrix}

	\SetAlgoLined
	\small
	\KwData{Simulation parameters (see Section \ref{sec:cwtg_input})}
	\KwResult{A Workload Trace}
	$serviceList$ = initializeAllServices()\;
	\ForEach{T between the startTime and the endTime}{
	    $livingServiceList$ = GetAllLivingServicesAtTime($serviceList$, $t$)\;
	    \ForEach{$livingService$ in $livingServiceList$}{
	        $vmQuantity$ = GetVmQuantityFromHorizontalElastFunc()\;
	        \textbf{GenerateVmsForService($livingService$, $vmQuantity$, $t$)}
	    }
	}
	print the simulated $serviceList$ and their VMs\;
\end{algorithm}

The generation of VM requests for a specific service at a discrete time $t$ is explained in Algorithm \ref{alg_vm_generation}. In general terms, the generation of VM requests is done by counting how many VM requests the service currently has, removing by setting and end time to randomly selected VM requests if the number of actual associated requests exceeds the number of intended VM requests that the service should have at the discrete time $t$ or creating new requests with a starting time $t$ if the number is actual number of VMs is smaller than the intended number, visible in the loops from line $1$ to $3$ and $4$ to $6$ respectively. Finally, the number of VMs that are on at the discrete time $t$ are iterated over and snapshots are added to them representing their state at said unit time $t$. The snapshot created contains the current instance type and the current resource utilization rates simulated using the vertical elasticity and the overbooking distribution functions.

\begin{algorithm}[!ht]
\caption{Generate VMs for a Service}
 \label{alg_vm_generation}
	\SetKwFunction{Union}{Union}
	\SetKwFunction{UpdatePlacementMatrix}{UpdatePlacementMatrix}

	\SetAlgoLined
	\small
	\KwData{Service, VmQuantity, CurrentTime}
	\While{getLivingVmCount(Service) \textgreater VmQuantity}{
	    choose a random VM from the $Service$ and mark it as finished\;
	}
	
	\While{getLivingVmCount(Service) \textless VmQuantity}{
	    create a $newVm$ and add it to the VM list of the $Service$\;
	}
	$livingVmList$ = GetAllLivingVmsAtTime($Service$, $CurrentTime$)\;
	\ForEach{livingVm in livingVmList}{
	    creates a snapshot of the state of time $CurrentTime$ of the $livingVm$\;
	}
\end{algorithm}

\section{Output Data}
\label{sec:output_data}
The output data of the Cloud Workload Trace Generator represents the workload of a dynamic environment. This workload is represented by comma separated values (CSV) file. The output of a simple test environment (See Table \ref{tab:input_example_cwtg}) with no Elasticity nor Overbooking will be used as an example. In the generated trace (See Table \ref{tab:output_example_cwtg}), three VMs can be identified, $V_{000}$ ($b_j=0$, $c_i$ and $V_j=0$), $V_{001}$ ($b_j=0$, $c_i=0$ and $V_j=1$) and $V_{002}$ ($b_j=0$, $c_i=0$ and $V_j=2$). Since horizontal elasticity is not considered all the VMs are associated to the same service - datacenter pair ($b_0$ - $c_0$) and they are created ($t=1$) and destroyed ($t=9$) all at the same time.

\begin{table}[ht]
	
	\caption{Workload example for the VMP problem considering a No Elasticity, No Overbooking environment}
	\label{tab:output_example_cwtg}
	
	\ra{1.3}
	\centering
	
	\resizebox{37em}{!} {
	    \begin{tabular}{@{} *{15}{c} @{}}
    		\toprule
            $t$ & $b_j$ & $c_j$ & $v_j$ & $cpu$ & $ram$ & $net$ & $u_{cpu}$ & $u_{ram}$ & $u_{net}$ & $r_{cpu}$ & $r_{ram}$ & $r_{net}$ & $t_{init}$ & $t_{end}$ \\
    		\midrule
    		0&    0&    0&    0&    6&    8&    450&    100&    100&    100&    0.065&    0.016&    0.179&    0&    1\\
            0&    0&    0&    1&    3&    75&   500&    100&    100&    100&    0.065&    0.016&    0.179&    0&    1\\
            0&    0&    0&    2&    6&    8&    450&    100&    100&    100&    0.065&    0.016&    0.179&    0&    1\\
            \midrule
            1&    0&    0&    0&    6&    8&    450&    100&    100&    100&    0.065&    0.016&    0.179&    0&    1\\
            1&    0&    0&    1&    3&    75&   500&    100&    100&    100&    0.065&    0.016&    0.179&    0&    1\\
            1&    0&    0&    2&    6&    8&    450&    100&    100&    100&    0.065&    0.016&    0.179&    0&    1\\
    		\bottomrule
    	\end{tabular}
	}
\end{table}

Furthermore, the resources requested by the VMs remain the same throughout the workload trace since vertical elasticity is not considered. If the vertical elasticity was included in the simulation parameters, the amount of resources allocated for a VM would change, this is represented by a change on the instance type associated to a VM at a specific time $t$.

The utilization of resources by the VMs is not dynamic and stays at a constant value of a 100\% following the exclusion of the overbooking uncertainty parameter for both server and network resources. In case the overbooking factor is considered, the utilization ratios will change over time.

\chapter{Part II - Evaluated Algorithms}
\label{chap:evaluated_algorithm_second_part}

In order to analyze the most suitable alternatives to solve the formulated VMP problem (see Chapter \ref{chap:problem_formulation_second_part}), an experimental evaluation of three scalarization methods for the iVMP and VMPr phases was performed.

The evaluated scalarization methods are the Weighted Sum method (WS) \cite{lopez2016LANC}, the distance to origin by the Euclidean method (ED) \cite{cha2007comprehensive} and the distance to origin by the Chebyshev method (CD) \cite{cha2007comprehensive}.

As part of the experiment, the Best Fit Decreasing (BFD) algorithm was used in the iVMP phase. a MA based on the one proposed by Ihara et al. in \cite{Ihara2015f} is considered for an offline resolution of the formulated VMP problem. Both algorithms were presented in Chapter \ref{chap:algorithms_first_part}.



The following is a list detailing the SMs considered in the experiments:

\section{Weighted Sum Method}
\label{subsec:weighted_sum}

\vspace{0.5cm}
\begin{equation}\label{eq:wsm}
	\begin{aligned}
		min f(x,t) = \sum_{i=1}^{q} \hat{f_i}(x,t) \times w_i
	\end{aligned}
\end{equation}

\vspace{2cm}
\noindent where:

\vspace{-2mm}
\begin{tabbing}
\hspace{3cm}\=\kill
	${f}(x,t)$:         \>Combined costs of the objective functions $f_i(x,t)$ at instant $t$;\\
	$\hat{f_i}(x,t)$:   \>Normalized cost of objective function $f_i(x,t)$ at instant $t$;\\
	$w_i$:              \>Weight of importance associated to $f_i(x,t)$;\\
	$q$:                \>Number of objective functions. In this case 4.
\end{tabbing}

In this work, the weight values $w_i$ associated to each objective function $f_i(x)$ are equivalent values in the range of $(0,1)$ and the sum of the weight values must be $1$. Considering the four objective functions presented in Section \ref{sec:objective_functions_second_part}, the weight value for each OF will be $0.25$.

\section{Minimum Distance to Origin: Euclidean Method}
\label{subsec:min_dist_euclidean}

\vspace{-2mm}
\begin{equation}\label{eq:initial_euclidean}
	\begin{aligned}
		min f(x,t) = \sqrt{\sum_{i=1}^{q} {\Big| \hat{f_i}(x,t) - \hat{f_{i_{opt}}}(x,t) \Big|}^2 }
	\end{aligned}
\end{equation}

Considering $\hat{f_{i_{otp}}}(x,t)$ to be the optimal value of the normalized objective function cost $\hat{f_i}(x,t)$ at a specific time $t$, then its value would be $0$ taking into account the minimization context of this article. Consequently, the eq. \eqref{eq:initial_euclidean} can be reduced to:

\vspace{-2mm}
\begin{equation}
    \label{eq:euclidean_method}
	\begin{aligned}
		min f(x,t) = \sqrt{\sum_{i=1}^{q} {\Big| \hat{f_i}(x) \Big| }^2 }
	\end{aligned}
\end{equation}

\vspace{-2mm}
\noindent where:

\vspace{-2mm}
\begin{tabbing}
\hspace{1.5cm}\=\kill
	${f}(x,t)$:         \>Combined costs of the objective functions $f_i(x,t)$ at instant $t$;\\
	$\hat{f_i}(x,t)$:   \>Normalized cost of objective function $f_i(x,t)$ at instant $t$; \\
	$q$:                \>Number of objective functions. In this case 4.
\end{tabbing}

\section{Minimum Distance to Origin: Chebyshev Method}
\label{subsec:min_dist_chebyshev}

\vspace{-2mm}
\begin{equation}
    \label{eq:initial_chebysev}
	\begin{aligned}
		min f(x,t) = max \bigg( \big|\hat{f_i}(x,t) - \hat{f_{i_{opt}}}(x,t)\big| \bigg) \quad \forall i \in Q
	\end{aligned}
\end{equation}

Considering $\hat{f_{i_{otp}}}(x,t)$ to be the optimal value of the normalized objective function cost $\hat{f_i}(x,t)$, then the equation \eqref{eq:initial_chebysev} can be reduced to:

\vspace{-2mm}
\begin{equation}
    \label{eq:chebysev_method}
	\begin{aligned}
		min f(x,t) = max \bigg( \big|\hat{f_i}(x,t)| \bigg) \quad \forall i \in Q
	\end{aligned}
\end{equation}

\vspace{-2mm}
\noindent where:

\vspace{-2mm}
\begin{tabbing}
\hspace{1.5cm}\=\kill
	${f}(x,t)$:         \>Combined costs of the objective functions $f_i(x,t)$ at instant $t$;\\
	$\hat{f_i}(x,t)$:   \>Normalized cost of objective function $f_i(x,t)$; \\
	$Q$:                \>Objective function set. In this case with a size of 4.
\end{tabbing}

\chapter{Part II - Experimental Results}
\label{chap:experimental_results_second_part}

\section{Experimental Environment}
\label{sec:experimental_env_part_two}

All evaluated algorithms were implemented using the Java programming language. The source code is available online\footnote{\url{https://github.com/DynamicVMP/dynamic-vmp-framework}}, as well as all the considered experimental inputs and outputs.
Experiments were performed on a GNU Linux System with an Intel(R) Xeon(R) E5530 at 2.40 GHz CPU and 16 GB of RAM.

Physical resources (matrix $H$ \eqref{matrix:H}) represent an heterogeneous IaaS cloud, considering four physical machine types (i) Small (S), (ii) Medium (M), (iii) Large (L) and (iv) Extra Large (XL), shown in Table \ref{tab:unique_pm}.

\begin{table}[!hb]

	\caption{Considered Physical Machine Types}
	\label{tab:unique_pm}
	
	\ra{1.3}
	\centering
	
	\resizebox{37em}{!} {
    \begin{tabular}{l*{4}{c}}
    		\toprule
    		& \multicolumn{4}{c}{\textbf{Resource}} \\
    		\cmidrule(lr){2-5}
    		\textbf{Type} & \textbf{CPU (ECU)} & \textbf{RAM (GB)} & \textbf{Network Capacity (Mbps)} & \textbf{\textit{pmax} (W)} \\
    		\midrule
    		\textit{Small (S)} & 32 & 128 & 1000 & 800 \\
    		\textit{Medium (M)} & 64 & 256 & 1000 & 1000  \\
    		\textit{Large (L)} & 256 & 512 & 1000 & 3000 \\
    		\textit{XLarge (XL)} & 512  & 1024 & 20000 & 5000 \\
    		\bottomrule
        \end{tabular}
    }
\end{table}

Taking into account that the considered VMP problem formulation is evaluated under uncertainty considering a scenario-based modeling approach, experimental workload traces were carefully designed for the performed experiments. The CWTG proposed in Chapter \ref{chap:cwtg} is implemented to produce workload traces that simulate the uncertainty factors required by the formulation.


This work defines two CPU load scenarios by limiting the number of PMs assigned to each datacenter, therefore using the same workload trace with a very large set of resources and a constrained set of resources in order to simulate load scenarios.
The low CPU load scenario considered 50 $S$, $M$ and $L$ PM types as well as 30 $XL$, while the high CPU load scenario considered 20 $S$ and $M$ PM types as well as 15 $L$ and 8 $XL$ PM types respectively.

To generate the traces utilized in the experiments, a combination of two different probability distribution functions were considered with different arguments each. A Uniform PDF and a Poisson PDF. Considering four parameters with two different distribution functions each, the total number of scenarios generated reaches 16; Since the distribution functions are not deterministic, the generator was executed with the same combination of parameters and arguments three times yielding a number of 48 scenarios. The different workload traces were considered and evaluated with the 2 defined CPU load scenarios previously described, totalling 96 different generated scenarios. Each experiment was executed ten times per scenario and the average results are presented in the next section. The following sections summarize the main findings in the performed experimental evaluation. All parameters considered for generating the experimental workload traces are presented in Table \ref{tab:inputs_cwtg}.

\begin{table}[!hb]
	\caption{Considered Data for Generated Experimental Workload Traces}
	\label{tab:inputs_cwtg}
	
    \ra{1.3}
	\centering
	
	\begin{tabular}{@{} ll @{}}
		\toprule
		\bfseries Parameter Description & \bfseries Input Data \\
		\midrule
		\multirow{2}{*}{1. Environment} & Overbooking (Server and Network resources)\\ 
							 			& Elasticity (Vertical and Horizontal) \\
        \\
		2. Workload Trace Duration [$t$] & 1000 \\ 
		3. Number of Cloud Datacenters & 1 \\
		4. Number of Cloud Services & 100 \\
		5. Number of VMs per Service & 10 \\
		6. Horizontal Elasticity & Uniform(0,10) \& Poisson($\lambda$=7) \\
  		7. Vertical Elasticity & Uniform(0,10) \& Poisson($\lambda$=5) \\
		8. Server Resources & Uniform(0,100) \& Poisson($\lambda$=70) \\
		9. Network Resources & Uniform(0,100) \& Poisson($\lambda$=70) \\
		\bottomrule
	\end{tabular}
\end{table}

\section{Experiment: Scalarization Methods}
\label{sec:experiments_part_two}

The main goal of this experiment is to answer the following question: \textit{is there any preferable scalarization method for solving the two-phase formulation of provider-oriented VMP problems?}

To answer the above mentioned question, each considered objective function is evaluated in a pure multi-objective context, as next described. Taking into account the findings described in \cite{amarillaevaluating}, this experiment was performed considering only BFD online heuristic with a protection factor of $\lambda_k=0.75$. Additionally to the average evaluation criteria presented in Equation \eqref{eq:avg}, each objective function cost was independently evaluated in a pure multi-objective context considering Pareto dominance and preference comparison method \cite{von2014survey}. Experiment results are summarized in Table \ref{tab:resultsExp3} and MFs can be enunciated as:

\begin{table}[!hb]

    \caption{Average objective function costs of the evaluated scalarization methods considering the BFD online heuristic with a protection factor of $\lambda_k=0.75$.}
	\label{tab:resultsExp3}
	
	\ra{1.3}
	\centering
	
	\resizebox{37em}{!} {
    	\begin{tabular}{@{} *{6}{l}*{2}{r} }
    		\toprule
    		& & & & & & \multicolumn{2}{c}{\textbf{CPU Load}}\\
    		\cmidrule{7-8}
    		\textbf{OF} & \textbf{Algorithm} & \textbf{Protection Factor} & \textbf{SM} & & & \textbf{Low} & \textbf{High} \\
            \midrule
    		\multirow{3}{*}{\textit {$f_1(x,t)$}} & \multirow{3}{*}{\textit {BFD}} & \multirow{3}{*}{\textit{$\lambda=0.75$}} & \textit{WS}  & & & 68246 & 41407 \\
    		& & & \textit{ED} & & & \cellcolor[HTML]{d3d3d3}\textbf{68026} & \cellcolor[HTML]{d3d3d3}\textbf{41144} \\
    		& & & \textit{CD} & & & 70571 & 43448 \\
    		\midrule
    		\multirow{3}{*}{\textit {$f_2(x,t)$}} & \multirow{3}{*}{\textit {BFD}} & \multirow{3}{*}{\textit{$\lambda=0.75$}} & \textit{WS} & & & \cellcolor[HTML]{d3d3d3}\textbf{287696} & 314213 \\
    		& & & \textit{ED} & & & 289247 & 315657 \\
    		& & & \textit{CD} & & & 288298 & \cellcolor[HTML]{d3d3d3}\textbf{307588} \\
    		\midrule
    		\multirow{3}{*}{\textbf {$f_3(x,t)$}} & \multirow{3}{*}{\textit {BFD}} & \multirow{3}{*}{\textit{$\lambda=0.75$}} & \textit{WS} & & & 0.539 & \cellcolor[HTML]{d3d3d3}\textbf{0.460} \\
    		& & & \textit{ED} & & & \cellcolor[HTML]{d3d3d3}\textbf{0.525} & 0.466 \\
    		& & & \textit{CD} & & & 0.529 & 0.472 \\
    		\bottomrule
        \end{tabular}
    }
\end{table}

\textbf{MF7:} \textit{All evaluated scalarization methods obtained Pareto optimal solutions, considering the Pareto dominance relation.}

When evaluating each of the obtained solutions with different scalarization method (i.e. WS, ED and CS) in a multi-objective context, all the alternatives obtained non-dominated solutions therefore the utilization of any of the evaluated scalarization methods is reasonably good. Consequently, a more fine-grained evaluation was performed by considering a preference relation between non-dominated solutions. A solution is defined as preferred to another when it is better in more objective functions \cite{lopez2016LANC}.

\textbf{MF8:} \textit{The minimum Euclidean distance to origin is preferable to the Weighted Sum and the Chebyshev distance to origin as the scalarization method of the proposed two-phase VMP problem.}

The Euclidean distance to origin can be singled out as the scalarization method that produces in average the best results for the three analyzed objective functions in both Low and High CPU load scenarios. This would suggest that using this method is recommended to achieve a good performance in all of the considered OFs of the problem's formulation.

\textbf{MF9:} \textit{The minimum Euclidean distance to origin is suggested to optimize the Power Consumption OF considering the results in the two CPU load scenarios analyzed.}

Experimental results also indicate that the minimum Euclidean distance obtained the best values of the power consumption objective function ($f_1(x,t)$) in both evaluated CPU load scenarios, as can be seen in Table \ref{tab:resultsExp3}.

\chapter{Part II - Conclusion and Future Work}
\label{chap:conclusions_and_future_works_second_part}

The second part of this work proposed an experimental evaluation of three scalarization methods developed (\textbf{Objective 4}, published in \cite{amarillaevaluating}) in order to find the most suitable to be part of the two-phase optimization scheme for VMP problems in federated-clouds considering many objective functions, elasticity and overbooking (\textbf{Objective 5}, published in \cite{amarillaevaluating}).

Due to the randomness of customer requests, the VMP problem were evaluated under uncertainty, considering different uncertain parameters. Said uncertain parameters were modeled through workload traces used as input of the formulation. To create the workload traces, a Cloud Workload Trace Generator was proposed and implemented (\textbf{Objective 3}, published in \cite{amarillaevaluating}).

The main findings indicate that all of the studied scalarization methods perform reasonably good when compared with each other given that all of them obtained non-dominated solutions considering a Pareto dominance relation.

A more detailed analysis shows that the Euclidean distance to origin is preferable over the other studied scalarization methods given that in average its utilization yields better solutions. Given the considered power consumption objective function, it can be said that the Euclidean distance to origin is recommended to ensure the best values of the objective function in any CPU load scenario.

Several future work were identified. First, a formulation of a VMP problem considering a dynamic set of PMs $H(t)$, to consider PM crashes, maintenance or even deployment of new generation hardware. Considering VMP formulations with more sophisticated cloud federation approaches is also left as a future work. Also, an experimental evaluation of alternative scalarization methods as part of the two-phase formulation is proposed as a future work.

In order to generate more realistic workload traces, other distribution functions can be implemented as part of the CWTG, as well as a mechanism to switch between distribution functions in the same workload trace generation. 

Additionally, an extended experimental evaluation of different parameters of the proposed VMP formulation should be considered. Experimenting with geo-distributed datacenters is left as future work. Finally, implementing the evaluated algorithms in real-world IaaS middlewares (e.g. OpenStack) in order to perform an evaluation in real-world cloud computing datacenters supporting real-world applications.

\clearpage
\lhead{\emph{References}} 
\addtotoc{References}
\btypeout{References}

\bibliographystyle{plain} 


\end{document}